\documentclass[review]{elsarticle}

\makeatletter
\def\ps@pprintTitle{%
	\let\@oddhead\@empty
	\let\@evenhead\@empty
	\def\@oddfoot{\centerline{\thepage}}%
	\let\@evenfoot\@oddfoot}
\makeatother

\usepackage[T1]{fontenc}
\usepackage[top=2cm, bottom=2cm, left=2cm, right=2cm]{geometry}
\usepackage{graphicx}%
\usepackage{epstopdf}
\usepackage{multirow}%
\usepackage{amsmath,amssymb,amsfonts}%
\usepackage{amsthm}%
\usepackage{mathrsfs}%
\usepackage[title]{appendix}%
\usepackage{xcolor}%
\usepackage{textcomp}%
\usepackage{manyfoot}%
\usepackage{booktabs}%
\usepackage{algorithm}%
\usepackage{algorithmicx}%
\usepackage{algpseudocode}%
\usepackage{listings}%
\usepackage{subfigure}%

\usepackage{amssymb}
\usepackage{amsmath}
\usepackage{amsthm}
\usepackage{xr}
\usepackage{lineno,hyperref}
\usepackage{cleveref}
\usepackage{epsfig}
\usepackage{arydshln}
\usepackage{float}
\usepackage{verbatim}
\usepackage{comment}
\newcommand{\comm}[1]{}
\usepackage{color}
\usepackage{xcolor}
\usepackage{booktabs}
\usepackage{threeparttable}
\usepackage{gensymb}
\usepackage{upgreek}

\newcommand{\bC}[0]{\mathbf{C}}

\newcommand{\bI}[0]{\mathbf{I}}

\newcommand{\bR}[0]{\mathbf{R}}

\newcommand{\bfi}[0]{\mathbf{\upphi}}
\newcommand{\bsig}[0]{\mathbf{\upsigma}}
\newcommand{\beps}[0]{\mathbf{\upvarepsilon}}
\newcommand{\br}[0]{\mathbf{r}}
\newcommand{\bs}[0]{\mathbf{s}}

\newcommand{\bfig}[0]{\bfi_g}

\newcommand{\GU}[0]{G_{\rm{U}}}
\newcommand{\GL}[0]{G_{\rm{L}}}
\newcommand{\Ki}[0]{\bar{K}^{\rm{iso}}}
\newcommand{\Gi}[0]{\bar{G}^{\rm{iso}}}
\newcommand{\Ei}[0]{\bar{E}^{\rm{iso}}}
\newcommand{\nui}[0]{\bar{\nu}^{\rm{iso}}}
\newcommand{\tC}[0]{\mathbb{C}}
\newcommand{\tS}[0]{\mathbb{S}}
\newcommand{\tA}[0]{\mathbb{A}}

\newcommand{\tP}[0]{\mathbb{P}}
\newcommand{\tI}[0]{\mathbb{I}}
\newcommand{\tCg}[0]{\tC_g}

\newcommand{\tCo}[0]{\tC_0}
\newcommand{\tCs}[0]{\tC_*}
\newcommand{\tCa}[0]{\bar{\tC}}
\newcommand{\tCi}[0]{\tCa^{\rm{iso}}}
\newcommand{\tCV}[0]{\tCa_{\rm{V}}}
\newcommand{\tCR}[0]{\tCa_{\rm{R}}}
\newcommand{\tCHU}[0]{\tCa_{\rm{*HU}}}
\newcommand{\tCHL}[0]{\tCa_{\rm{*HL}}}
\newcommand{\tCHS}[0]{\tCa_{\rm{HS}}}
\newcommand{\tCHSU}[0]{\tCa_{\rm{HSU}}}
\newcommand{\tCHSL}[0]{\tCa_{\rm{HSL}}}
\newcommand{\tCSC}[0]{\tCa_{\rm{SC}}}
\newcommand{\tID}[0]{\tI_{\rm{D}}}
\newcommand{\tIP}[0]{\tI_{\rm{P}}}

\usepackage{lineno}

\begin{document}

\begin{frontmatter}

\title{Multiscale evaluation of the elastic response of polycrystalline NiAl using molecular statics, EBSD-based finite element and analytical homogenization methods}

\author[addr1]{Szymon~Nosewicz\corref{cor1}}
\ead{snosew@ippt.pan.pl}

\author[addr1]{Marcin~Ma\'zdziarz}
\ead{mmazdz@ippt.pan.pl}

\author[addr1]{Pawe\l{}~Ho\l{}obut}
\ead{pholob@ippt.pan.pl}

\author[addr2]{Witold~Chromi\'{n}ski}
\ead{witold.chrominski@pw.edu.pl}

\author[addr1]{Katarzyna~Kowalczyk-Gajewska}
\ead{kkowalcz@ippt.pan.pl}

\author[addr1]{Grzegorz~Jurczak}
\ead{gjurcz@ippt.pan.pl}

\cortext[cor1]{Corresponding author}
\address[addr1]{Institute of Fundamental Technological Research, Polish Academy of Sciences, \\ Pawi\'nskiego 5B, 02-106 Warsaw, Poland}
\address[addr2]{Faculty of Materials Science and Engineering, Warsaw University of Technology\\
	Wo\l{}oska 141, 02-507 Warsaw, Poland}
 
\begin{abstract}
The presented work proposes a multiscale framework for the numerical and analytical determination of the elastic constants of the intermetallic nickel aluminide (NiAl) at the microscopic scale. Molecular dynamics (MD) calculations have been employed to evaluate the stiffness tensors of a single NiAl monocrystalline body with cubic Pm$\bar{3}$m symmetry. The atomistic results were transferred to the upper scale as an input to micromechanical models. Here, finite element modeling (FEM) and analytical homogenization methods were used to simulate the elastic behavior of polycrystalline NiAl, with special attention to grain anisotropy and its morphology. The microstructural features of the heterogeneous body were evaluated experimentally using electron backscatter diffraction (EBSD) and subsequently used to generate a 3D polycrystalline geometric model that preserves realistic distributions of grain orientations and grain sizes. The reconstructed NiAl polycrystal was subjected to uniaxial tensile test simulations using FEM to evaluate the effective Young's modulus ($E$) and Poisson's ratio ($\nu$). The numerical results were confronted with the experimental values of $E$ and~$\nu$. With these model assumptions, a very good agreement between numerical and experimental results was achieved. Finally, the experimental results were compared with classical analytical estimates. Among the analytical approaches, the Hashin–Shtrikman upper bound and self-consistent approximation reproduced the experimental Young's modulus with minor deviations.
\end{abstract}


\end{frontmatter}

	
    \newpage
    \section{Introduction}
    \label{sec:Int}
    

The effective mechanical response of polycrystalline materials arises from interactions among grains, grain boundaries, crystallographic orientations, porosity, and various microstructural defects distributed across multiple length scales \cite{patala_understanding_2019}. Unlike homogeneous solids, polycrystalline bodies exhibit pronounced local heterogeneity due to the elastic anisotropy of individual grains and the complexity of grain-boundary networks, making the prediction of effective properties significantly more demanding \cite{kube_elastic_2016}. This issue is particularly important in intermetallic materials such as nickel aluminide (NiAl), where {\color{black} the pronounced elastic anisotropy makes its deformation response orientation-dependent, highlighting the importance of crystallographic texture when evaluating the effective stiffness \cite{Frommeyer2010}. Resolving this anisotropic response provides a basis for analyzing load partitioning and local stress concentrations in NiAl-based composites, where phase arrangement and crack-microstructure interactions influence fracture resistance \cite{Gabel2021}. In NiAl bond coats used to protect nickel-based superalloys, the dependence of Young's modulus, residual stresses, and fracture toughness on local composition and thermal exposure further demonstrates the engineering relevance of accurate elastic characterization \cite{Webler2012}. 
The combination of low density, high melting temperature, high elastic modulus, and excellent oxidation resistance makes NiAl attractive for advanced engineering applications, motivating the development of reliable models to predict its mechanical response \cite{holobut_modeling_2026}.}

The elastic response of polycrystalline materials is commonly predicted using analytical homogenization, numerical simulations, atomistic-to-continuum multiscale schemes, and, more recently, data-driven surrogate models. Classical analytical approaches such as Voigt, Reuss, Voigt–Reuss–Hill, Hashin–Shtrikman or Hershey provide fast estimates of effective elastic constants, but they usually rely on strong assumptions concerning strain or stress uniformity, simplified grain interaction, random texture, or idealized aggregate morphology \cite{stefaniuk_voigt-reuss_2023}. More advanced analytical approaches, such as self-consistent schemes and Mori–Tanaka models \cite{kowalczyk-gajewska_elastic_2020}, attempt to account for grain shape and local heterogeneity. Although these methods are computationally efficient, they generally rely on simplified assumptions regarding grain morphology, orientation distributions, and stress transfer mechanisms. 

The numerical modeling of polycrystalline materials has been applied and explored in hundreds of papers using molecular dynamics \cite{lin_effect_2024}, discrete element modeling \cite{nouri_approach_2013}, phase field modeling \cite{darban_experimental_2022}, peridynamics \cite{liu_peridynamic_2023}, and finally finite element modeling \cite{quey_large-scale_2011}. However, it remains challenging due to the inherently multiscale nature of deformation and stress transfer mechanisms. At the atomic scale, the bonding character and crystal symmetry govern the intrinsic elastic properties of monocrystalline grains and grain boundaries \cite{mazdziarz_atomistic_2024}, while at the microscale, grain morphology, orientation distribution, and grain interactions determine local stress concentrations and deformation patterns \cite{karafiatova_fitting_2023}. These mechanisms collectively define the effective elastic and mechanical response of the bulk material at the macroscale; however, traditional continuum approaches based on homogeneous constitutive laws and simplified isotropic assumptions often neglect the influence of grain orientations and grain boundary characteristics, leading to inaccurate predictions of local stress fields and effective properties \cite{han_limitations_2008}. 

Multiscale modeling has emerged as one of the most promising approaches for investigating heterogeneous materials with complex internal structures \cite{song_multiscale_2024}. In such frameworks, lower-scale simulations provide constitutive information for higher-scale models, enabling a physically consistent description of the material response across multiple length scales \cite{fish_mesoscopic_2021}. Atomistic methods, particularly molecular dynamics (MD), enable direct investigation of interatomic interactions and evaluation of elastic constants, stress tensors, defect evolution, and grain-boundary properties \cite{huang_multiscale_2026}, while these atomistically determined parameters can subsequently be incorporated into continuum-scale numerical methods, such as the finite element method (FEM), for analyzing realistic microstructures at engineering length scales \cite{yamazaki_multiscale_2024}.

Most existing numerical studies employ statistically generated or idealized microstructures instead of experimentally reconstructed geometries, which may limit their ability to accurately capture realistic grain morphology, grain-size distribution, crystallographic texture, and local heterogeneity \cite{zheng_modeling_2022}. This limitation also extends to many recent machine-learning and physics-informed surrogate models, which are typically trained using synthetic periodic microstructures \cite{monteiro_fernandes_physics-informed_2025}. The use of experimentally reconstructed microstructures has become increasingly important for improving the predictive capability of multiscale models. Among the available characterization techniques, X-ray micro-computed tomography (micro-CT) enables the reconstruction of three-dimensional microstructures, whereas electron backscatter diffraction (EBSD) provides detailed information on grain morphology and crystallographic orientations, making it particularly suitable for simulations of anisotropic polycrystalline materials. Micro-CT enables three-dimensional visualization of internal structures and is particularly effective for analyzing porosity, cracks, and inclusions in heterogeneous materials \cite{liu_modeling_2025, li_deep_2025}. However, its application to polycrystalline reconstruction is limited by insufficient spatial resolution to accurately characterize nanoscale and microscale features, especially in fine-grained materials. Moreover, micro-CT does not directly provide crystallographic orientation information, which is essential for modeling anisotropic grain behavior. In contrast, EBSD offers detailed information on grain morphology, crystallographic orientation/texture, and grain-boundary distributions at significantly higher spatial resolution \cite{doddapaneni_advancements_2025}. As a result, EBSD-based reconstruction techniques have become highly attractive for generating realistic polycrystalline models for numerical simulations using FEM \cite{sun_microscale_2018,zhang_effects_2022, brands_computational_2016}. In particular, such methodologies can explicitly account for grain anisotropy and orientation-dependent stiffness, which play a critical role in intermetallic compounds and other strongly anisotropic materials. Furthermore, FEM-based analyses enable the direct comparison between numerical predictions and experimentally measured macroscopic properties, thereby providing a powerful validation framework for multiscale approaches.

Despite considerable progress in multiscale modeling and microstructure reconstruction techniques, accurately predicting the effective elastic properties of polycrystalline intermetallic materials remains an open challenge. In particular, the integration of atomistic information with realistic FEM-based polycrystalline reconstructions requires careful consideration of scale transitions, grain morphology representation, and anisotropic constitutive behavior. Therefore, the present work proposes a comprehensive multiscale numerical–analytical-experimental framework for determining the elastic constants of polycrystalline NiAl by combining molecular dynamics simulations, EBSD-based microstructure reconstruction, and finite element modeling. Special attention is devoted to the realistic representation of grain anisotropy and morphology, as well as to the assessment of classical analytical homogenization approaches against experimentally validated numerical predictions.

\newpage

	\section{Fundamentals of the methodology}
	\label{sec:Fotmam}
	\subsection{Atomistic modeling of NiAl crystals}
	\label{sec:Amosc}

{\color{black} The atomistic calculations were performed to characterize the anisotropic elastic response of NiAl within the adopted interatomic description and
to provide consistent, reproducible constitutive input for the subsequent
micromechanical analysis.} The elastic properties necessary for the performance of micromechanical models were evaluated within molecular dynamics simulations for B2-NiAl at the single-crystal level. Simulations of three uniaxial and three shear deformations were performed in order to evaluate the complete stiffness tensor $\tC$. Owing to the pronounced cubic elastic anisotropy of NiAl \cite{holobut_modeling_2026}, the elastic response of individual grains strongly depends on their crystallographic orientation with respect to the loading direction. 

\begin{figure}[H]
    \centering
    \includegraphics[width=1\linewidth]{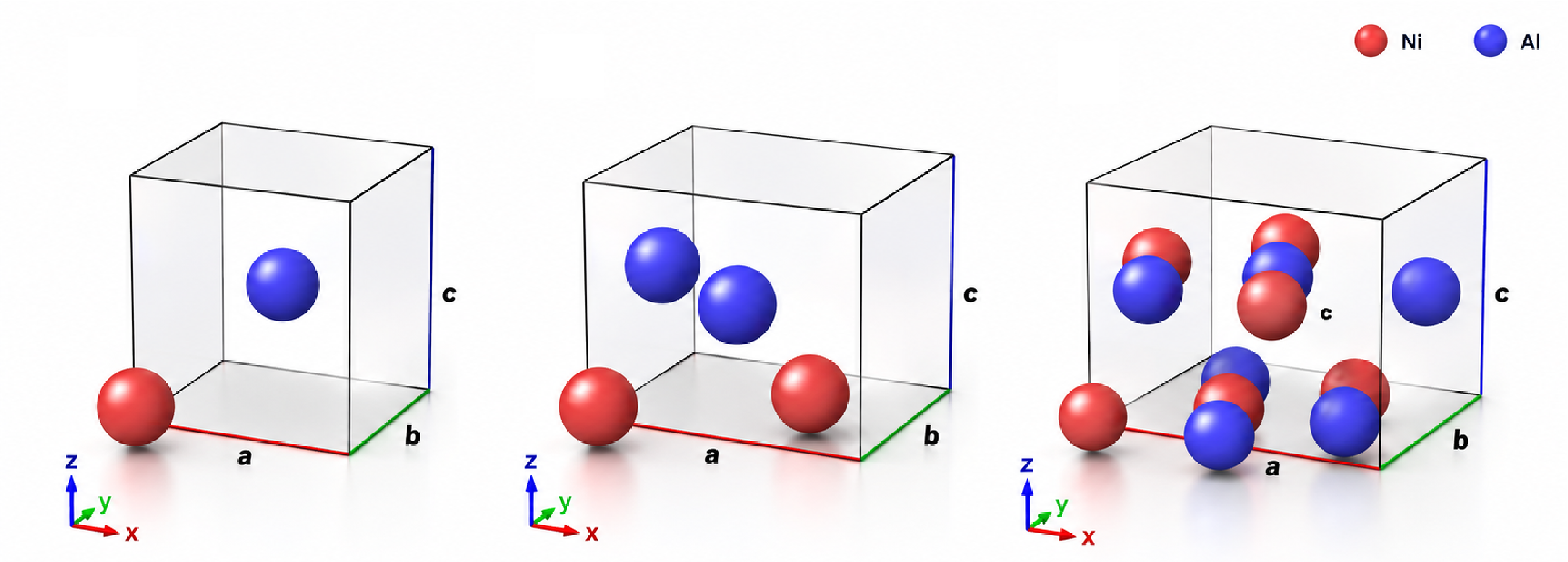} 
    { (a)} \hspace{5cm} { (b)} \hspace{5cm} { (c)} \\
    \caption{Atomistic models of NiAl basic cell: a) \textbf{x}=[100] \textbf{y}=[010] \textbf{z}=[001], b) \textbf{x}=[110] \textbf{y}=[-110] \textbf{z}=[001], c) \textbf{x}=[111] \textbf{y}=[-1-12] \textbf{z}=[1-10].}
    \label{fig:NiAl_geo}
\end{figure}

    All atomistic calculations were carried out in accordance with the methodology described in \cite{mazdziarz_atomistic_2024, Mazdziarz2025am}, i.e., with the molecular dynamics (MD) approach, the Large-scale Atomic/Molecular Massively Parallel Simulator (\href{https://www.lammps.org/}{LAMMPS}) \cite{LAMMPS}, periodic boundary conditions (PBC) in all directions, and the embedded-atom method (EAM) interatomic potential \cite{Purja2009}. {\color{black} Although an earlier potential co-developed by Mishin \cite{Mishin2002} reproduces the experimental single-crystal elastic constants of B2-NiAl more closely, the EAM2009 potential adopted here offers broad transferability across the binary Ni-Al system, including the reproduction of several properties not included in its fitting database.} The \href{https://atomsk.univ-lille.fr/}{Atomsk} \cite{Hirel2015212} software was used to generate nickel-aluminium (B2-NiAl) computational cells for three different orientations of lattice vectors: the \textbf{x}=[100], \textbf{y}=[010] and \textbf{z}=[001] directions, see Fig.~\ref{fig:NiAl_geo}\,a), the \textbf{x}=[110], \textbf{y}=[-110] and \textbf{z}=[001] directions, see Fig.~\ref{fig:NiAl_geo}\,b), and the \textbf{x}=[111], \textbf{y}=[-1-12] and \textbf{z}=[1-10] directions, see Fig.~\ref{fig:NiAl_geo}\,c). The \href{https://www.ovito.org/}{OVITO} \cite{Ovito2010} software was used for the visualization.
	The stiffness tensors for all pre-relaxed structures were obtained by using the stress--strain method with a maximum strain amplitude of 10$^{-4}$ \cite{LAMMPS}.

\comm{\begin{figure}[H] 
	\centering
	\begin{tabular}{ccc}
		{\bf a)} & {\bf b)} & {\bf c)} \\
		\includegraphics[width=0.310\linewidth]{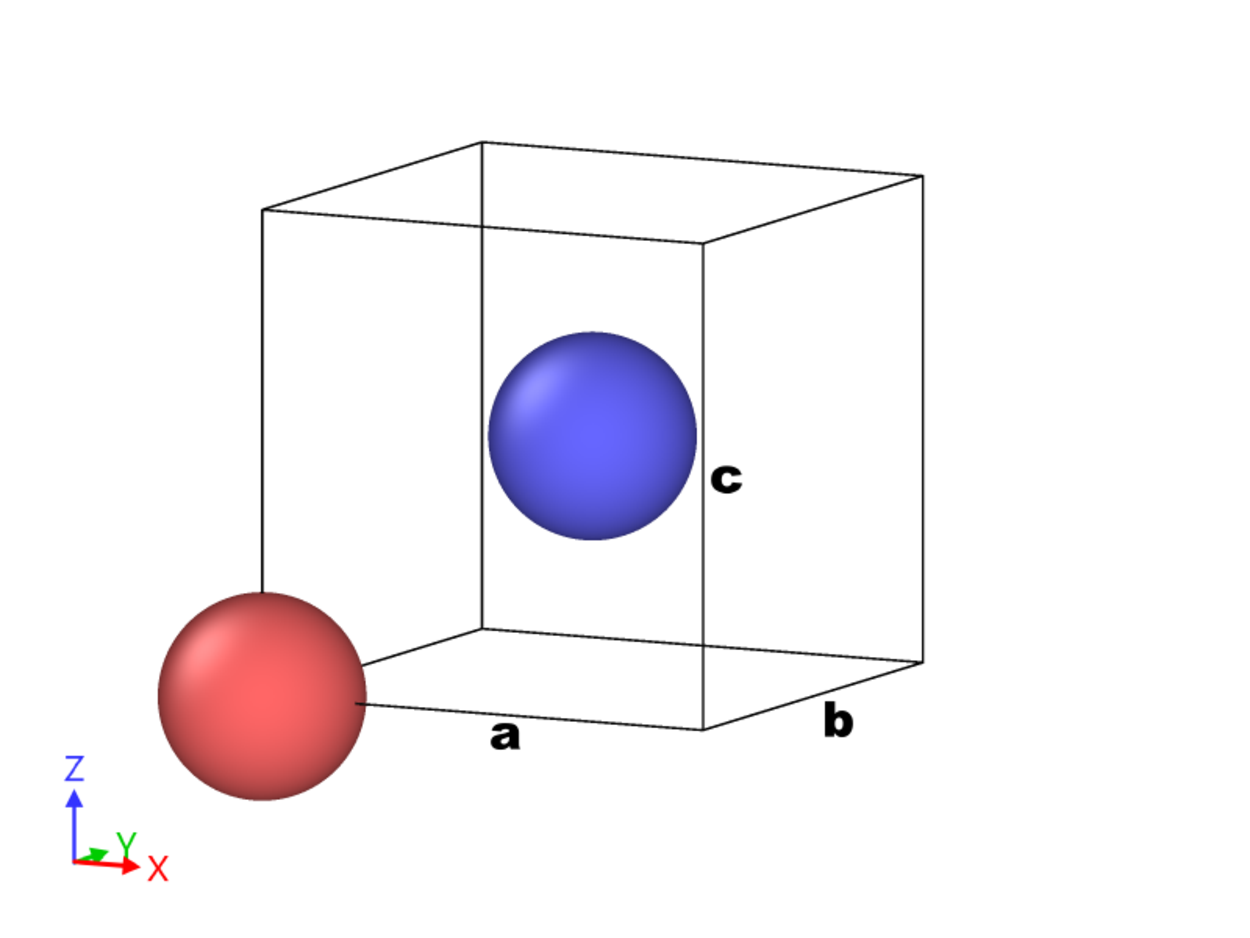} &
		\includegraphics[width=0.310\linewidth]{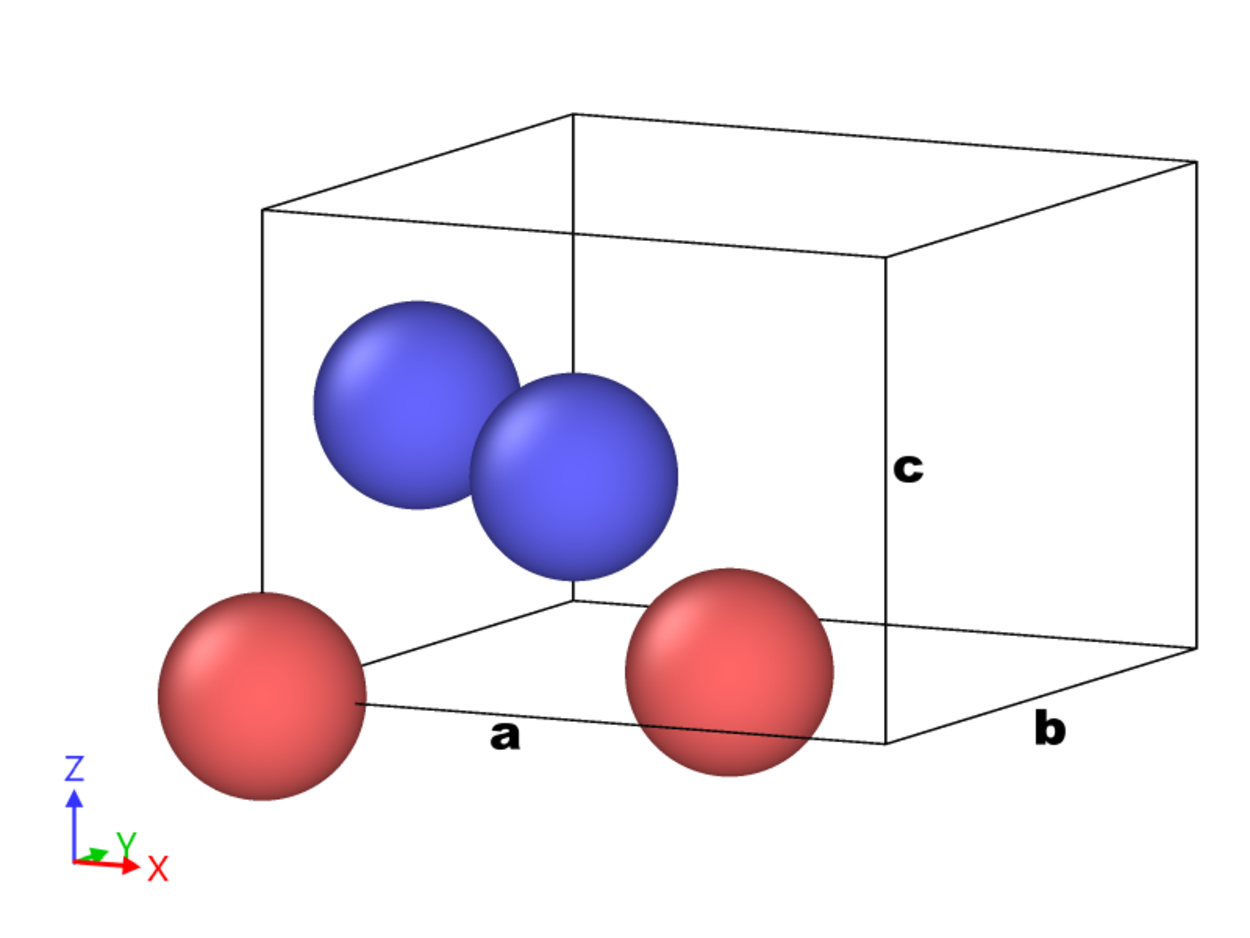} &
		\includegraphics[width=0.320\linewidth]{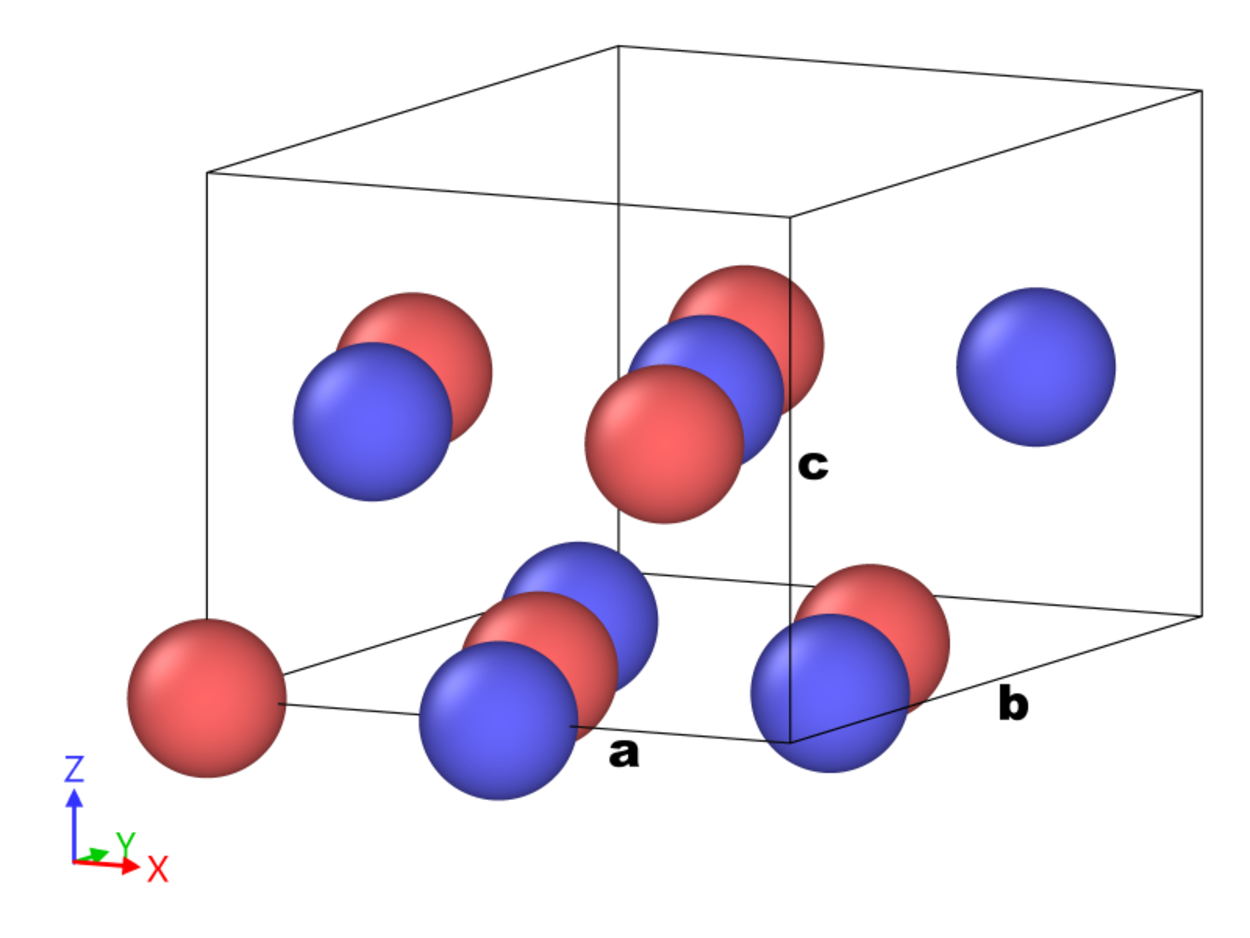} \\
	\end{tabular}
	\caption{NiAl: a) basic cell \mathbf{x}=[100] \mathbf{y}=[010] \mathbf{z}=[001], b) basic cell x=[110] y=[-110] z=[001], c) basic cell x=[111] y=[-1-12] z=[1-10]\comm{, d) polycrystal amorphous} (The red and blue atoms represent Ni and Al, respectively).}
	\label{fig:NiAl_geo3}
\end{figure}}

    \subsection{EBSD reconstruction methodology}
    \label{sec:EBSD}

\begin{figure}[!ht]
    \centering
    \includegraphics[width=1\linewidth]{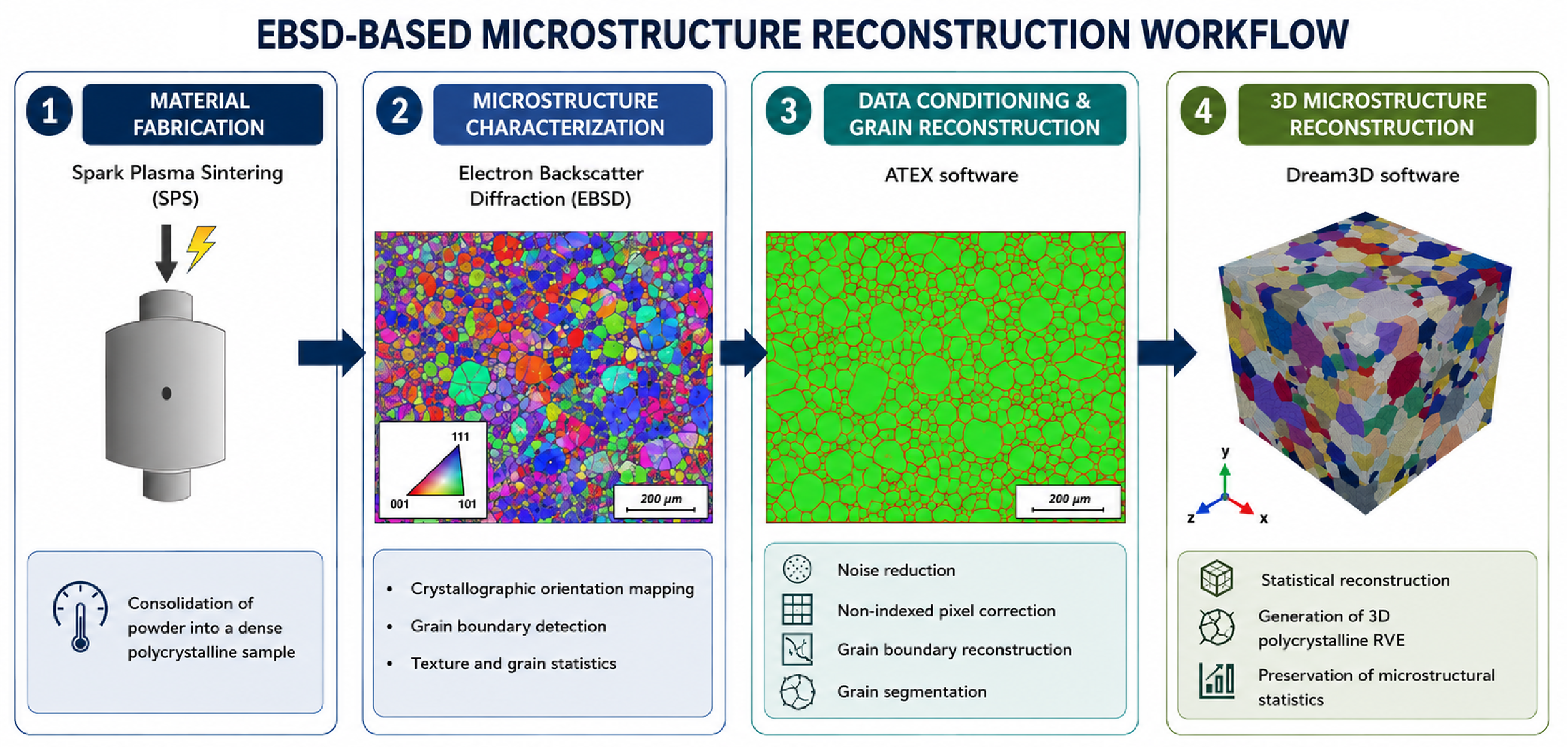}
    \caption{The overview of the EBSD-based reconstruction workflow.}
    \label{fig:GenerationWorkflow}
\end{figure}

Since accurate prediction of material behavior depends on the realistic representation of microstructural features, constitutive response, and loading conditions, there is an increasing demand for robust computational tools capable of generating and simulating high-fidelity microstructures for structure–property analyses \cite{ge_3d_2026}. This work focuses on generating statistically equivalent synthetic microstructures using statistical characterization data from EBSD measurements. The proposed reconstruction approach aims to provide computational modeling with representative microstructures that reproduce the key statistical features of the actual polycrystalline material. The overview has been presented in Fig.~\ref{fig:GenerationWorkflow}.

As a representative material for the experimental and numerical investigations, the intermetallic NiAl material was selected. Intermetallic NiAl powder (Goodfellow, purity 99.9\%, gas-atomized) was consolidated using the spark plasma sintering (SPS) technique, in which a direct electric current simultaneously heats the powder compact and the die, while a uniaxial mechanical load promotes densification \cite{abedi_reactive_2023}. A key advantage of SPS, associated with the short processing time, is the suppression of excessive grain growth, which is particularly important for preserving fine-grained or nanostructured microstructures \cite{bregiroux_spark_2021}. The NiAl powder was heated to the target sintering temperature of 1300$^o$C and maintained at an external pressure of 30\,MPa for 10\,min to achieve nearly complete densification \cite{nosewicz_combined_2022}.

Next, two-dimensional EBSD measurements were performed on mechanically polished surfaces, followed by short-duration Ar+ ion polishing immediately prior to analysis to improve signal quality. Particular attention was devoted to surface preparation, as minimizing unresolved points due to poor signal from grain interiors was essential to the reliability of the study. The measurements were conducted using a field-emission scanning electron microscope (Hitachi SU70) equipped with an HKL Nordlys EBSD detector, operating at an acceleration voltage of 20\,kV.

The experimentally obtained EBSD data were subsequently processed using the ATEX software package, a non-commercial tool dedicated to the analysis of electron backscatter diffraction and X-ray diffraction data \cite{beausir_ATEX_2017}. To reduce the fraction of non-indexed (unresolved) pixels and improve the continuity of the orientation maps, several post-processing procedures, including data cleaning, interpolation and smoothing algorithms, were applied \cite{korznikova_microstructural_2012}. Since unresolved pixels may also occur within grain interiors due to local signal limitations, additional correction procedures based on noise reduction and spike removal were implemented to reconstruct missing regions \cite{beausir_ATEX_2017}. Finally, the processed orientation data enabled the generation of a pixel-based representation of the microstructure, in which neighboring pixels with similar crystallographic orientations were grouped to define individual grain objects.

Finally, the 2D EBSD data were transferred to Dream3D software, which was used for a wide range of applications, from basic EBSD analysis to advanced 3D microstructure reconstruction \cite{groeber_dream3d_2014}. Dream3D software works as a sequence of modules for microstructure generation, including: equivalent ellipsoidal grain generator, constrained grain packer, seed point generator, constrained Voronoi tessellation tool and crystallographic orientation assignment \cite{groeber_framework_2008}. The collection of these modules and the experimental EBSD characterization processes constitutes an automated methodology for simulating representative polycrystalline microstructures, which has been applied in several very recent papers \cite{luo_microplasticity_2026, kashyap_tribological_2026, anto_size_2026}. 

The synthetic microstructure generation methodology has previously been validated and shown to reproduce the key morphological and crystallographic characteristics of real polycrystalline materials with good statistical agreement \cite{groeber_framework_2008}. In the present study, two main quantitative microstructural descriptors are considered: the grain-size distribution and the crystallographic orientation distribution, {\color{black} with individual grain orientations represented by Euler angles, $\bfig=(\phi_1,\phi_2,\phi_3)_g$}. The synthetic microstructure was considered statistically representative when these output distributions closely matched the corresponding distributions obtained from the 2D EBSD measurements. The resulting three-dimensional microstructure was subsequently converted into a finite element mesh, serving as the preprocessor for the FE model. The Dream3D-generated geometry was discretized using the Parallelized Polycrystal Mesher developed by Cerrone et al. \cite{zhang_eigenstrain_2015}.

	\subsection{Finite element modeling of polycrystalline sample}
    \label{sec:FEM}

\begin{figure}[!ht]
    \centering
    \includegraphics[width=0.75\linewidth]{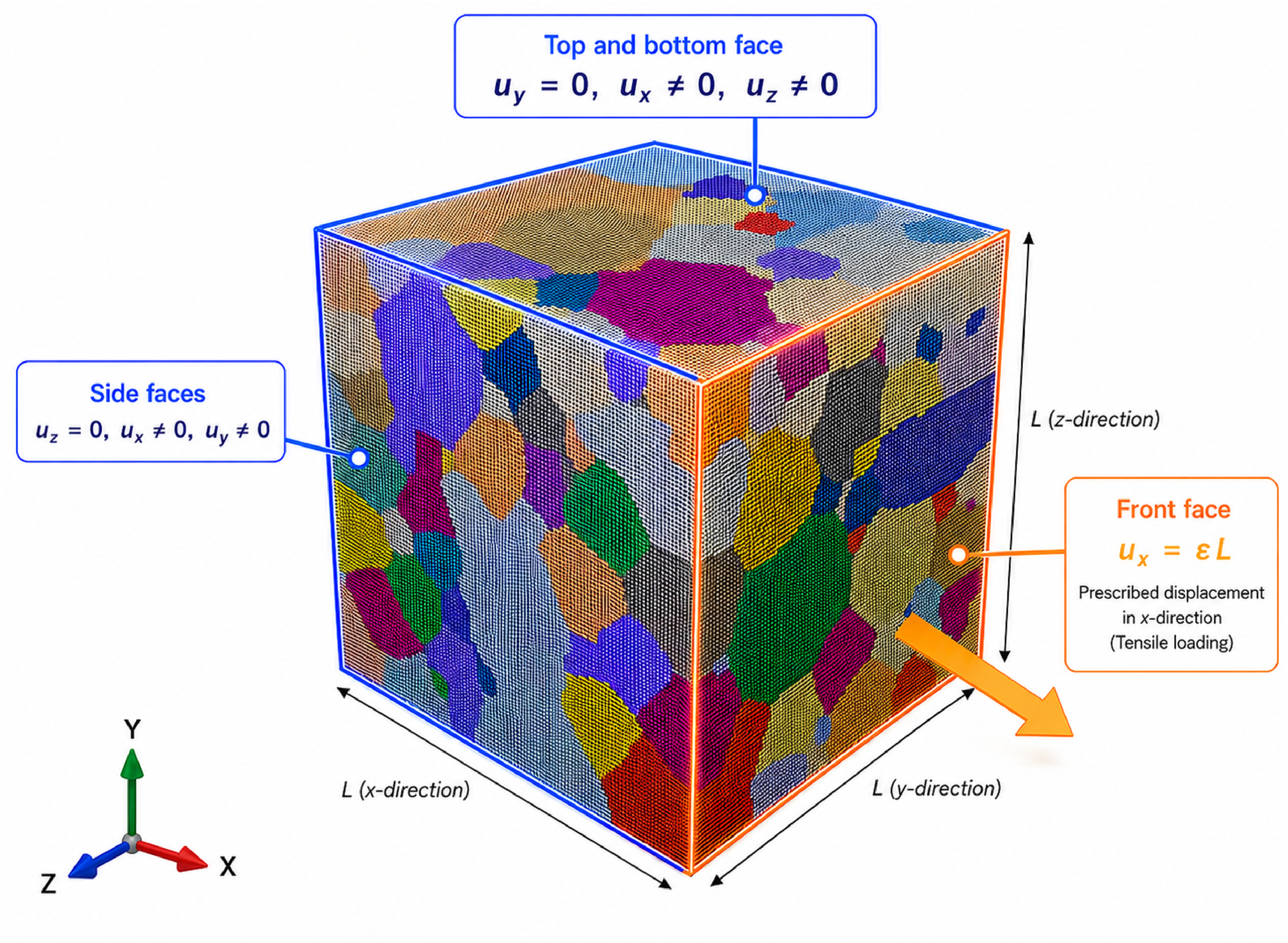}
		\caption{Boundary conditions applied to the EBSD-based representative volume element (RVE) used in finite element simulations.}
		\label{fig:BC}
\end{figure}

Finite element simulations were performed to determine the effective elastic properties of polycrystalline NiAl by explicitly incorporating (i) the atomistically determined single-crystal stiffness tensor of NiAl to define the elastic behavior of individual grains, and (ii) the experimentally reconstructed grain morphology and crystallographic orientations obtained from EBSD measurements. The boundary-value problem of linear elasticity was solved using the finite element method implemented in ABAQUS \cite{smith_abaqusstandard_2009}. Assuming linear elastic material behavior, and using standard summation over repeated indices, the constitutive relation is expressed as:

{\begin{equation}
	\centering
    \sigma_{ij} = C_{ijkl} \varepsilon_{kl}\,,
    \label{Elast1}
\end{equation}}

\noindent where the local stress tensor $\bsig$ with components $\sigma_{ij}$ and strain tensor $\beps$ with components $\varepsilon_{ij}$ are proportional, with the proportionality coefficients given by the anisotropic stiffness tensor $\tC$ with components $C_{ijkl}$ in the local crystal frame. Within the single-grain NiAl crystal, cubic symmetry is represented by 3 independent components of the stiffness tensor. However, arbitrary rotation of the grain's local crystallographic orientation, as provided by EBSD measurements and defined by Euler angles $\bfig$, changes its representation in the global reference frame. The components of the stiffness tensor generally change, and previously zero components become nonzero. The transformation rule is as follows:
{\begin{equation}
	\centering
    C_{ijkl}' = R_{im} R_{jn} R_{ko} R_{lp} C_{mnop}\,,
    \label{Elast2}
\end{equation}}

\noindent where $R_{ij}$ are components of the orthogonal transformation matrix $\bR$ from the principal coordinate system of the grain to the global coordinate system of the sample and $C_{ijkl}'$ are components of the stiffness tensor of the grain expressed in the global coordinate system \cite{Ostrowska2010}. {\color{black} The rotated stiffness tensor was specified for each grain using the fully anisotropic elastic material definition in ABAQUS; this accommodates its general representation in the global coordinate system while preserving the underlying cubic symmetry.} Dream3D software, based on EBSD images of polycrystalline NiAl (see Fig.~\ref{fig:GenerationWorkflow}), creates a discrete grain shape within the polycrystalline NiAl sample, as shown in Fig.~\ref{fig:BC}. A cubic sample of a given $L \times L \times L$ size (with a finite element mesh of up to $150 \times 150 \times 150$ C3D8 brick elements \cite{Mazdziarz2010}) comprises individual grains maintaining a $1 \times 1 \times 1\,\mu\mathrm{m}$ element resolution.

To determine the effective isotropic elastic properties of polycrystalline NiAl - Young's modulus ($E$) and Poisson's ratio ($\nu$), a uniaxial deformation was applied through kinematic boundary constraints. Specifically, displacement-controlled (kinematic) boundary conditions were applied by constraining the displacement component normal to each external surface, except for the loaded ("front") face, where a prescribed displacement was imposed, see Fig.~\ref{fig:BC}. The corresponding reaction forces were then used to determine the macroscopic stress response. 

Assuming an equivalent isotropic elastic material characterized by the Lam\'{e} constants: $\lambda$ and $\mu$, the non-zero stiffness coefficients can be expressed (in Voigt notation \cite{Nye1985}, using $6\times6$ matrix $\bC$ representation of the stiffness tensor ${\tC}$, with components $C_{IJ}$) as:
\begin{equation}
    C_{11} = \lambda + 2{\mu}, \quad C_{12} = \lambda, \quad C_{44} = {\mu}.
    \label{Elast3}
\end{equation}
By calculating normal stresses averaged over surfaces, we may calculate effective isotropic elastic parameters of the sample, as follows (here, for deformation along the \textbf{x} direction, i.e., $\varepsilon_{x} \neq 0$): 

\begin{equation}
    \begin{aligned}
    E = \frac{\mu (3\lambda + 2\mu)}{\lambda + \mu}, &\quad \nu = \frac{\lambda}{2(\lambda + \mu)}, \\
    \lambda' = \frac{\sigma_{y}}{\varepsilon_{x}}, \quad \mu' = \frac{\sigma_{x} - \sigma_{y}}{2\varepsilon_{x}}, &\quad \lambda'' = \frac{\sigma_{z}}{\varepsilon_{x}}, \quad \mu'' = \frac{\sigma_{x} - \sigma_{z}}{2\varepsilon_{x}}.
    \end{aligned}
    \label{Elast4}
\end{equation}

\noindent This approach results in two equivalent sets of effective elastic properties ($E'$, $\nu'$, and $E''$, $\nu''$, as the forces measured on two surfaces perpendicular to the displacement direction may differ) multiplied by 3 different deformation tests (i.e., along the \textbf{x}, \textbf{y}, and \textbf{z} axes).

To evaluate the representativeness of the reconstructed polycrystalline RVEs, the coefficient of variation ($CoV$) and the elastic anisotropy index ($A$) were introduced for both Young's modulus ($CoV(E)$, $A_E$) and Poisson's ratio ($CoV(\nu)$, $A_\nu$) \cite{anto_size_2024}. The parameter $CoV(E)$ quantifies the scatter of the effective Young's modulus among multiple evaluations of a given RVE size and is defined as:

\begin{equation}
CoV(E)=\frac{SD(E)}{\overline{E}}\times100\%,
\label{eq:cov}
\end{equation}

\noindent where, $SD(E)$ is the standard deviation while $\overline{E}$ is the mean effective Young's modulus $E$ of the polycrystalline sample. The elastic anisotropy index $A_E$ characterizes the directional dependence of the effective elastic response obtained for loading along the global coordinate axes and is expressed as

\begin{equation}
A_E=
\frac{\max(E_x,E_y,E_z)-\min(E_x,E_y,E_z)}
{\left(E_x+E_y+E_z\right)/3}
\times100\%,
\label{eq:anisotropy}
\end{equation}

\noindent where $E_x$, $E_y$, and $E_z$ denote the effective Young's moduli determined for loading along the global directions. As a statistically representative polycrystal is expected to exhibit isotropic behavior at the macroscopic scale, lower values of both $CoV(E)$ and $A_E$ indicate improved representativeness and convergence toward an isotropic effective elastic response.

    \subsection{Analytical models of effective elastic properties of polycrystalline materials}
    \label{sec:Cm}

The analytical mean-field homogenization methods employed in this study use the same multiscale framework as the finite element simulations (Sec.~\ref{sec:FEM}). The elastic stiffness tensor of single-crystalline NiAl, determined from molecular dynamics calculations, serves as the constitutive input, while the grain orientations and volume fractions obtained from the EBSD-based reconstruction define the polycrystalline aggregate. The analytical approaches considered include the Voigt and Reuss bounds, the upper and lower Hashin--Shtrikman bounds, and the self-consistent estimate, whose formulations are summarized below based on \cite{kowalczyk2018atomistic}. It should be noted that these estimates, in contrast to FEM, do not account for the inhomogeneity of the strain and stress fields inside individual grains.

We consider a polycrystal occupying a volume $V$ composed of $N_G$ grains, with $V_g$ being the volume of grain $g$, $g=1,2,\ldots,N_G$. For any quantity $\chi$ which may vary between grains but is constant inside each grain, we denote its value in grain $g$ by $\chi_g$, and use the averaging operation over all grains:
\begin{equation}\label{eq:average}
\langle \chi_g \rangle = \frac{1}{V}\sum_{g=1}^{N_G} \chi_g V_g.
\end{equation}
The adopted procedure for estimating Young's modulus, $E$, and Poisson's ratio, $\nu$, for a given finite polycrystal is as follows. First, the polycrystal's effective stiffness tensor $\tCa$ is calculated; in general, it is anisotropic. Then, the best isotropic approximation of $\tCa$ with respect to the Log-Euclidean metric is computed, denoted by $\tCi$. Finally, the isotropic $E$ and $\nu$ are obtained from $\tCi$. The above steps are described in more detail below. 

The analytical estimates of the effective stiffness tensor of an NiAl polycrystal are computed as follows \cite{kowalczyk2018atomistic}:
\begin{itemize}
\item The Voigt ($\rm{V}$) and Reuss ($\rm{R}$) estimates, which result from the assumption of uniform strain or uniform stress, respectively, in the entire polycrystal \cite{hill1952elastic} and which are independent of the shapes of grains, are:
\begin{equation}\label{eq:vbound}
\tCV=\left<\tCg\right>, 
\end{equation}
\begin{equation}\label{eq:rbound}
\tCR=\left<\tCg^{-1}\right>^{-1},
\end{equation}
where $\tCg=\tC(\bfig)$ is the stiffness tensor of grain $g$, expressed in the global coordinate system, whose crystal axes are rotated with respect to the global ones by the Euler angles $\bfig=(\phi_1,\phi_2,\phi_3)_g$.
	
\item The estimates resulting from the Hashin-Shtrikman ($\rm{HS}$) variational principle \cite{hashin1962some,hashin1962variational} can be presented as:
\begin{equation}\label{eq:hsbounds}
\tCHS=\left<(\tCg+\tCs(\tCo))^{-1}\right>^{-1}-\tCs(\tCo)\,,
\end{equation}
where $\tCs(\tCo)$ is the Hill tensor \cite{hill1965continuum} depending on the stiffness tensor $\tCo$ of a comparison material, assumed here to be isotropic, and the shape of the equivalent inclusion, assumed here to be spherical. The upper Hashin-Shtrikman bound, $\tCHSU$, and the lower bound, $\tCHSL$, are obtained by using bounding values of $\tCo$ selected in such a way as to ensure positive or negative definiteness, respectively, of the tensor $\tCo-\tC(\bfi)$ for any $\bfi$ \cite{walpole1981elastic,berryman2005bounds,kowalczyk2012estimation}. The bounds take the form:
\begin{equation}\label{eq:hsubound}
\tCHSU=\left<(\tCg+\tCHU)^{-1}\right>^{-1}-\tCHU\,,
\end{equation}
\begin{equation}\label{eq:hslbound}
\tCHSL=\left<(\tCg+\tCHL)^{-1}\right>^{-1}-\tCHL\,,
\end{equation}
where the respective Hill tensors are
\begin{equation}\label{eq:hillu}
\tCHU = 4 \GU \tIP + f(\GU,K) \tID,
\end{equation}
\begin{equation}\label{eq:hilll}
\tCHL = 4 \GL \tIP + f(\GL,K) \tID,
\end{equation}
and $\tIP$ (orthogonal projector to the hydrostatic subspace) and $\tID$ (orthogonal projector to the deviatoric subspace) have components $I_{ijkl}^{\rm P}=\frac{1}{3}\delta_{ij}\delta_{kl}$ and $I_{ijkl}^{\rm D}=\frac{1}{2}(\delta_{ik}\delta_{jl}+\delta_{il}\delta_{jk})-I_{ijkl}^{\rm P}$, respectively, in any orthonormal basis, $f(G,K)=G(8G+9K)/(6G+3K)$ (with $G=\GU$ in Eq.~\ref{eq:hillu} and $G=\GL$ in Eq.~\ref{eq:hilll}), $3K=C_{11}+2C_{12}$, $2G_1=C_{11}-C_{12}$, $G_2=C_{44}$ (with $C_{11}$, $C_{12}$ and $C_{44}$ being the three stiffness parameters of a cubic material in the crystal coordinate system), $\GU=\max(G_1,G_2)$ and $\GL=\min(G_1,G_2)$.

\item The self-consistent estimate ($\rm{SC}$), obtained by setting the strain in each grain $g$ to the strain in a spherical inclusion of stiffness $\tCg$ embedded in an infinite medium with the effective properties to be found \citep{kroner1958berechnung,hill1965continuum}, is given by the equation:
\begin{equation}\label{eq:scestimate}
\tCSC = \left<(\tCg + \tCs(\tCSC))^{-1}\right>^{-1} - \tCs(\tCSC)\,.
\end{equation}	
The above formula, utilizing the Eshelby result \cite{eshelby1957determination}, is implicit and must be solved iteratively.

\end{itemize}
 
Once an analytical estimate of the effective stiffness tensor $\tCa$ is found, its closest isotropic approximation $\tCi$ with respect to the Log-Euclidean metric is computed as proposed in \citep{moakher2006closest}. Below, the sign $:$ denotes full contraction of two tensors of the same dimensions and ${\rm{Log}\tA}=\sum_{L}(\log\lambda_L)\tP_L$, where $\lambda_L$ are the eigenvalues and $\tP_L$ the corresponding eigenprojectors obtained by spectral decomposition of a positive-definite symmetric fourth-order tensor $\tA$ \citep{kowalczyk2009bounds}. The isotropic approximation of $\tCa$ has the form
\begin{equation}\label{eq:ciso}
\tCi = 3 \Ki \tIP + 2 \Gi \tID \,,
\end{equation}
where the isotropic bulk modulus $\Ki$ and isotropic shear modulus $\Gi$ are given by
\begin{equation}\label{eq:kgiso}
\Ki = \frac{1}{3} \exp\left( \rm{Log}\tCa : \tIP \right)\quad {\rm and} \quad \Gi = \frac{1}{2} \exp\left(\frac{1}{5} \rm{Log}\tCa : \tID \right).
\end{equation}
The values of $\Ki$ and $\Gi$ are translated into the corresponding values of isotropic Young's modulus, $E=\Ei$, and Poisson's ratio, $\nu=\nui$, using standard conversion formulas
\begin{equation}\label{eq:kgtoen}
\Ei = \frac{9 \Ki \Gi}{3\Ki + \Gi} \quad {\rm and} \quad \nui = \frac{3 \Ki - 2 \Gi}{6 \Ki + 2 \Gi} \,.
\end{equation}
The anisotropy of $\tCa$ is quantified using the following factor:
\begin{equation}\label{eq:zeta}
\zeta = \frac{\|\rm{Log}\tCa - \rm{Log}\tCi\|}{\|\rm{Log}\tCa\|}\geq 0\,,
\end{equation}
where $\|\tA\|=\sqrt{\tA:\tA}$. 

It should be mentioned that all the estimates discussed in this section, when computed for polycrystals of local cubic symmetry like NiAl, give the same value of $\Ki$ equal to the $K$ of a single crystal.

	\section{Results}
	\label{sec:Res}
	\subsection{Molecular dynamics simulations} 
	\label{sec:Mds}
   
   As a result of the pre-relaxation of the analyzed structures, the following lattice constants were obtained: $a=b=c=2.93${\,\AA}, when oriented in the \textbf{x}=[100], \textbf{y}=[010] and \textbf{z}=[001] directions (Fig.~\ref{fig:NiAl_geo}\,a), $a=b=4.14${\,\AA} and $c=2.93${\,\AA}, when oriented in the \textbf{x}=[110], \textbf{y}=[-1-10] and \textbf{z}=[001] directions (Fig.~\ref{fig:NiAl_geo}\,b) and $a=5.075${\,\AA}, $b=7.177${\,\AA} and $c=4.14${\,\AA} when oriented in the \textbf{x}=[111], \textbf{y}=[-1-12] and \textbf{z}=[1-10] directions (Fig.~\ref{fig:NiAl_geo}\,c). The results obtained are consistent with those produced by rotating the initial crystal. Moreover, the determined lattice constant $a$, corresponding to the crystal orientation defined by \textbf{x}=[100], \textbf{y}=[010], and \textbf{z}=[001], shows satisfactory agreement with theoretical calculations available in the literature \cite{fu_elastic_2013, Mishin2002, Papanicolaou2003, Faraoun2005}.
   
   Subsequently, the corresponding stiffness tensors were determined and, in Voigt notation, are given by: 
   \comm{Nickel-aluminium (B2-NiAl) alloy crystallizes in the cubic Pm$\bar{3}$m space group, where the crystallographic axes of the crystal lattice are oriented in the $X$=[100], $Y$=[010] and $Z$=[001] directions, see Fig.~\ref{fig:NiAl_geo}\,a), with lattice constants $a=b=c=2.93${\,\AA}, when oriented in the x=[110], y=[-110] and z=[001] directions, see Fig.~\ref{fig:NiAl_geo}\,b), $a=b=4.14${\,\AA} and $c=2.93${\,\AA}, when oriented in the x=[111], y=[-1-12] and z=[1-10] directions, see Fig.~\ref{fig:NiAl_geo}\,c), $a=5.075${\,\AA}, $b=7.177${\,\AA} and $c=4.14${\,\AA}. \href{https://www.ovito.org/}{OVITO} \cite{Ovito2010} program}
    
    \begin{itemize}
    	\item {\hyperref[fig:NiAl_geo]{NiAl}} oriented \textbf{x}=[100] \textbf{y}=[010] \textbf{z}=[001]
    	{\begin{equation}
    			\centering
    			\left[C_{IJ}\right]\rightarrow\left[
    			\begin{array}{cccccc}
    				190.87 & 142.91 & 142.91 & 0. & 0. & 0. \\
    				142.91 & 190.87 & 142.91 & 0. & 0. & 0. \\
    				142.91 & 142.91 & 190.87 & 0. & 0. & 0. \\
    				0. & 0. & 0. & 121.49 & 0. & 0. \\
    				0. & 0. & 0. & 0. & 121.49 & 0. \\
    				0. & 0. & 0. & 0. & 0. & 121.49 \\
    			\end{array}
    			\right]\,[\text{GPa}].
    			\label{tab:NiAl1}
    	\end{equation}}
    	
    	\item {\hyperref[fig:NiAl_geo]{NiAl}} oriented \textbf{x}=[110] \textbf{y}=[-110] \textbf{z}=[001]
    	{\begin{equation}
    			\centering
    			\left[C_{IJ}\right]\rightarrow\left[
    			\begin{array}{cccccc}
    				288.37 & 45.40 & 142.91 & 0. & 0. & 0. \\
    				45.40 & 288.37 & 142.91 & 0. & 0. & 0. \\
    				142.91 & 142.91 & 190.87 & 0. & 0. & 0. \\
    				0. & 0. & 0. & 121.49 & 0. & 0. \\
    				0. & 0. & 0. & 0. & 121.49 & 0. \\
    				0. & 0. & 0. & 0. & 0. & 23.98 \\
    			\end{array}
    			\right]\,[\text{GPa}].
    			\label{tab:NiAl2}
    	\end{equation}}
    	
    	\item {\hyperref[fig:NiAl_geo]{NiAl}} oriented \textbf{x}=[111] \textbf{y}=[-1-12] \textbf{z}=[1-10]
    	{\begin{equation}
    			\centering
    			\left[C_{IJ}\right]\rightarrow\left[
    			\begin{array}{cccccc}
    				320.88 & 77.91 & 77.91 & 0. & 0. & 0. \\
    				77.91 & 288.38 & 110.41 & 0. & 0. & -45.96 \\
    				77.91 & 110.41 & 288.37 & 0. & 0. & 45.96 \\
    				0. & 0. & 0. & 88.98 & 45.96 & 0. \\
    				0. & 0. & 0. & 45.96 & 56.48 & 0. \\
    				0. & -45.96 & 45.96 & 0. & 0. & 56.48 \\
    			\end{array}
    			\right]\,[\text{GPa}].
    			\label{tab:NiAl3}
    	\end{equation}}
    \end{itemize}
	
For all 3 orientations, the Kelvin moduli, i.e., the eigenvalues of the stiffness tensor written in orthonormal notation \cite{Mazdziarz2025am}, are identical and are as follows $\kappa_i$=(476.684, 242.971, 242.971, 242.971, 47.96, 47.96), $i$=1,\dots,6. Since the Kelvin moduli are invariants of the stiffness tensor, their behavior provides strong evidence that the atomistic calculations were performed correctly. It should be mentioned that the Kelvin moduli can also be written as: $\kappa_1=3K$, $\kappa_2=\kappa_3=\kappa_4=2G_2$ and $\kappa_5=\kappa_6=2G_1$, where $K$, $G_1$ and $G_2$ are the bulk modulus and two shear moduli of NiAl, respectively. The calculated anisotropy of crystalline NiAl measured by the Zener ratio is
\begin{equation}\label{eq:zener}
A_Z = \frac{G_2}{G_1} = \frac{242.971}{47.96} = 5.07 \,.
\end{equation} 

\begin{figure}[!ht]
    \centering
    \subfigure[]{\includegraphics[width=0.45\textwidth]{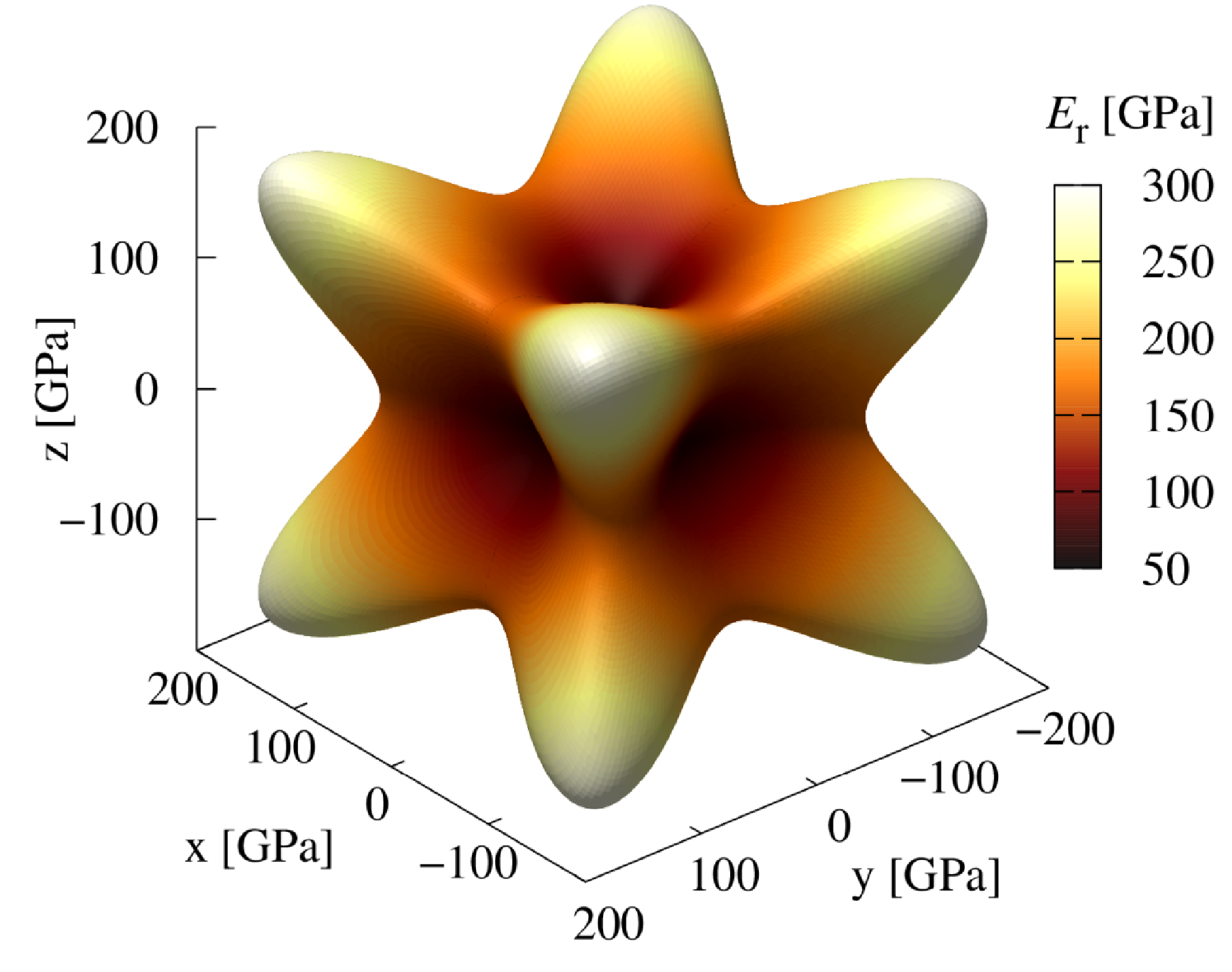}} \hspace{0.02\textwidth}
    \subfigure[]{\includegraphics[width=0.45\textwidth]{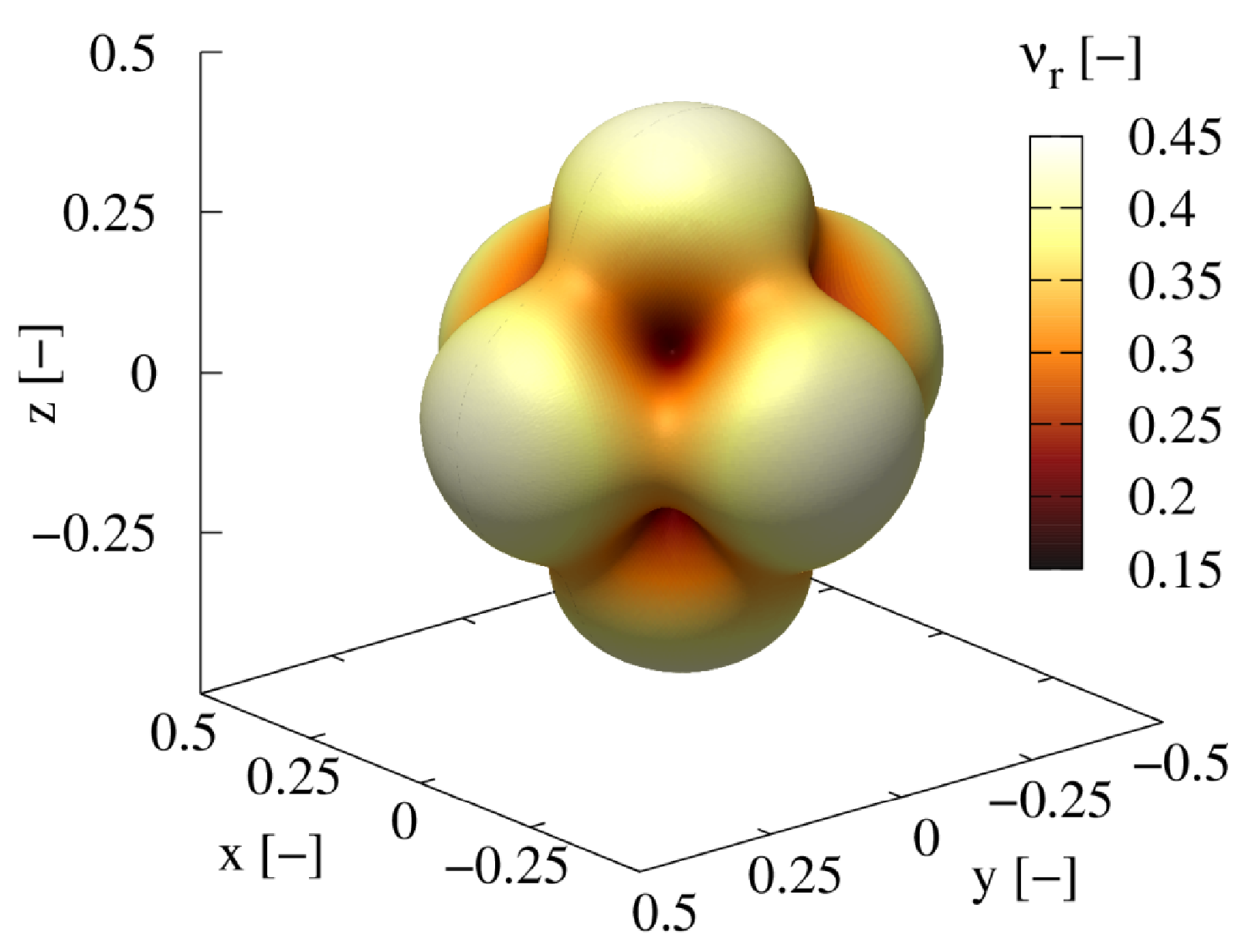}} \vspace{-0.2cm}
	\caption{Polar plots of the directional Young's modulus (a) and directional average Poisson's ratio (b). The distance from $\bf{0}$ to a given plot point is equal to the value of the plotted quantity in the direction of the plot point.}
	\label{fig:EP}
\end{figure}

Based on the evaluated atomistic results (Eqs.~\ref{tab:NiAl1}), directional Young's modulus and Poisson's ratio were computed (Fig.~\ref{fig:EP}) as follows. Young's modulus in the direction $\br$, where $\|\br\|=1$, is given by
\begin{equation}\label{eq:diryoung}
E_{\br} = \frac{\sigma_{\br}}{\varepsilon_{\br}} = \frac{1}{\br^2 \cdot \tS \cdot \br^2} \,,
\end{equation}
where $\sigma_{\br}$ is the magnitude of a uniaxial stress applied to the crystal in the direction $\br$, $\varepsilon_{\br}$ is the resulting strain along $\br$, $\tS=\tC^{-1}$ is the compliance tensor of the crystal, the sign $\cdot$ denotes contraction over two indices, and $\br^2=\br\otimes\br$ with $\otimes$ denoting the outer product. Poisson's ratio, in turn, depends on two vectors: the vector $\br$ along which uniaxial stress is applied to the crystal and the vector $\bs$, perpendicular to $\br$, along which the resulting lateral strain ($\varepsilon_{\bs}$) is measured. In general, the lateral deformation is different for different $\bs$. Poisson's ratio relative to directions $\br$ and $\bs$ is given by
\begin{equation}\label{eq:dirdirpoisson}
\nu_{\br\bs} = - \frac{\varepsilon_{\bs}}{\varepsilon_{\br}} = - \frac{\br^2 \cdot \tS \cdot \bs^2}{\br^2 \cdot \tS \cdot \br^2} \,.
\end{equation}
The average value of Poisson's ratio over all $\bs\perp\br$ is
\begin{equation}\label{eq:dirpoisson}
\nu_{\br} = \frac{1}{2\pi} \int_0^{2\pi} \nu_{\br\bs(\varphi)} {\rm{d}}\varphi = - \frac{1}{2} \frac{\br^2 \cdot \tS \cdot (\bI - \br^2)}{\br^2 \cdot \tS \cdot \br^2} \,,
\end{equation}
where $\bs(\varphi)$ covers all directions in the plane perpendicular to $\br$.

The graphical polar representation of $E$ (Fig.~\ref{fig:EP}a) reveals a pronounced elastic anisotropy associated with the cubic symmetry of the B2-NiAl crystal structure, characterized by maximum stiffness of 290.45~GPa along the $\langle111\rangle$ directions and minimum stiffness of 68.49~GPa along the $\langle100\rangle$ directions \cite{holobut_modeling_2026}. The shape of the Young's modulus surface reflects the underlying crystallographic symmetry and originates from the anisotropic arrangement of atoms within the crystal lattice (Fig.~\ref{fig:NiAl_geo}). Specifically, the $\langle111\rangle$ directions correspond to densely packed alternating Ni and Al atomic chains with shorter interatomic distances, resulting in stronger interatomic interactions and increased resistance to elastic deformation \cite{fu_elastic_2013}. In contrast, the $\langle100\rangle$ directions are characterized by larger atomic spacing and less efficient load-bearing atomic configurations, leading to lower stiffness values. 

The directional average Poisson’s ratio shown in Fig.~\ref{fig:EP}b exhibits a substantial degree of anisotropy, as evidenced by the pronounced lobes and depressions of the corresponding surface. The maximum value of average Poisson's ratio is 0.428 along the $\langle100\rangle$ directions, while the minimum value is 0.195 along the $\langle111\rangle$ directions. The extreme values of non-averaged Poisson's ratio are found for $\br$=[110], with the maximum value of 1.003 reached for $\bs$=[001] and the minimum of -0.340 for $\bs$=[1-10], and equivalent configurations. These results demonstrate that the elastic response of NiAl is intrinsically governed by the orientation-dependent distribution of atomic bonds, suggesting the need to determine the actual microstructure of polycrystalline aggregates, in which grain orientation significantly influences stress distribution and deformation behavior.

    \subsection{Generation of EBSD-based polycrystalline sample}
    \label{sec:GoEBSD}

\begin{figure}[!ht] 
		\centering
		\begin{tabular}{cc}
			\subfigure[]{\includegraphics[width=0.46\linewidth]{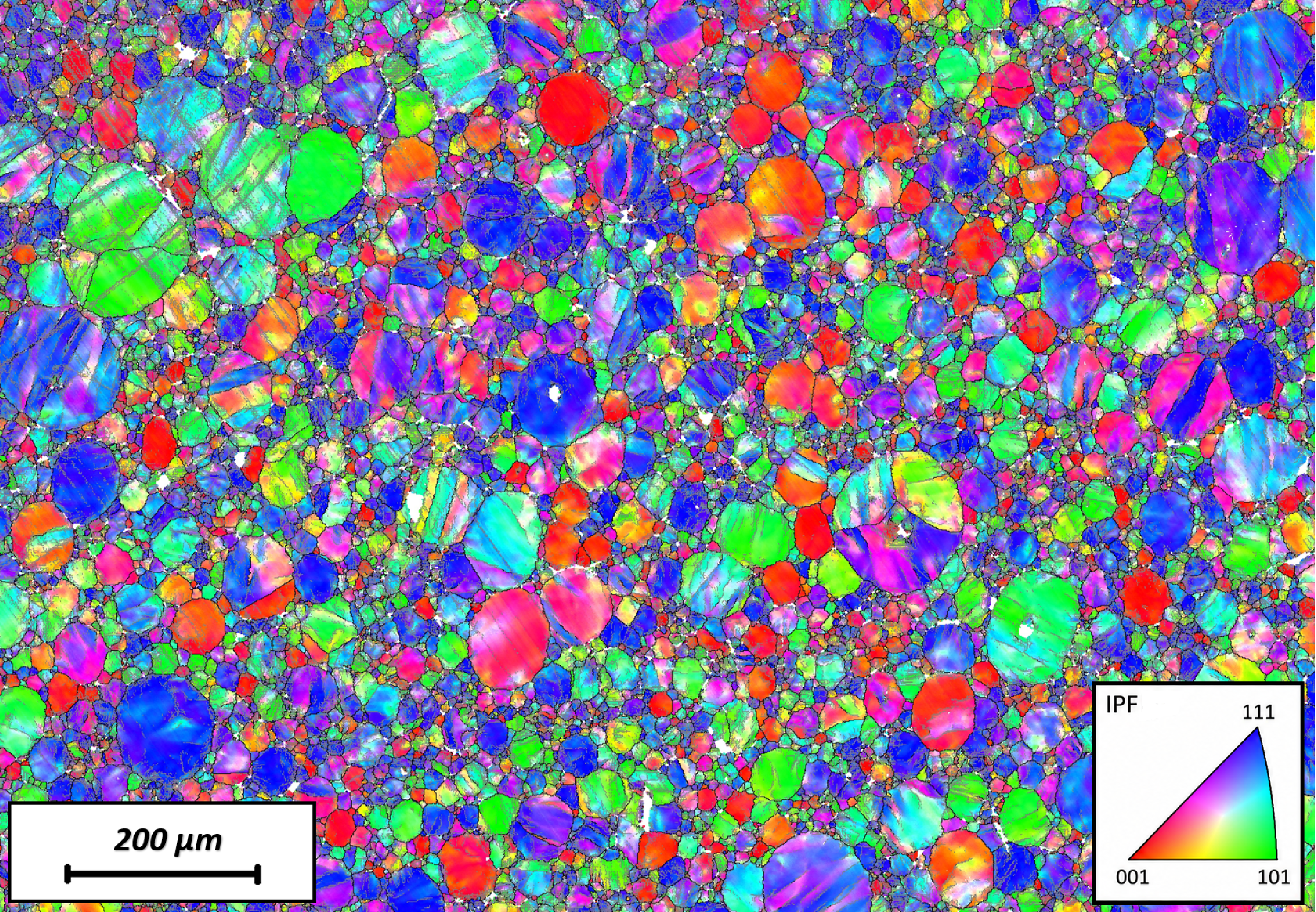}}
			\subfigure[]{\includegraphics[width=0.5030\linewidth]{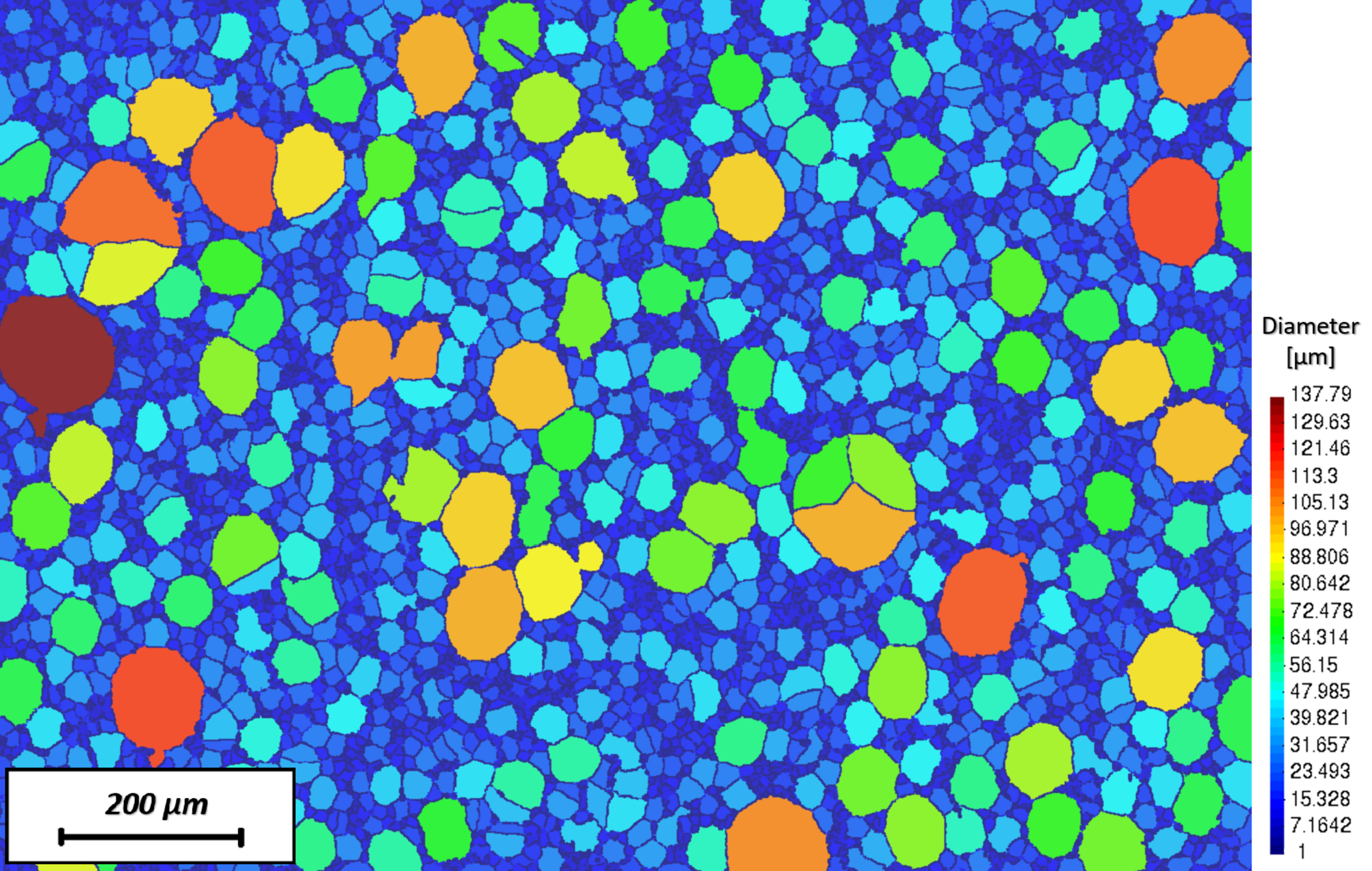}} \vspace{-0.2cm}
		\end{tabular}
		\caption{EBSD images of NiAl polycrystalline sample: a) as-received IPF mapping, b) corresponding smoothed and segmented map with a spatial grain size distribution.}
		\label{fig:EBSDmaps}
	\end{figure}

\begin{figure}[H]
    \centering
    \includegraphics[width=1\linewidth]{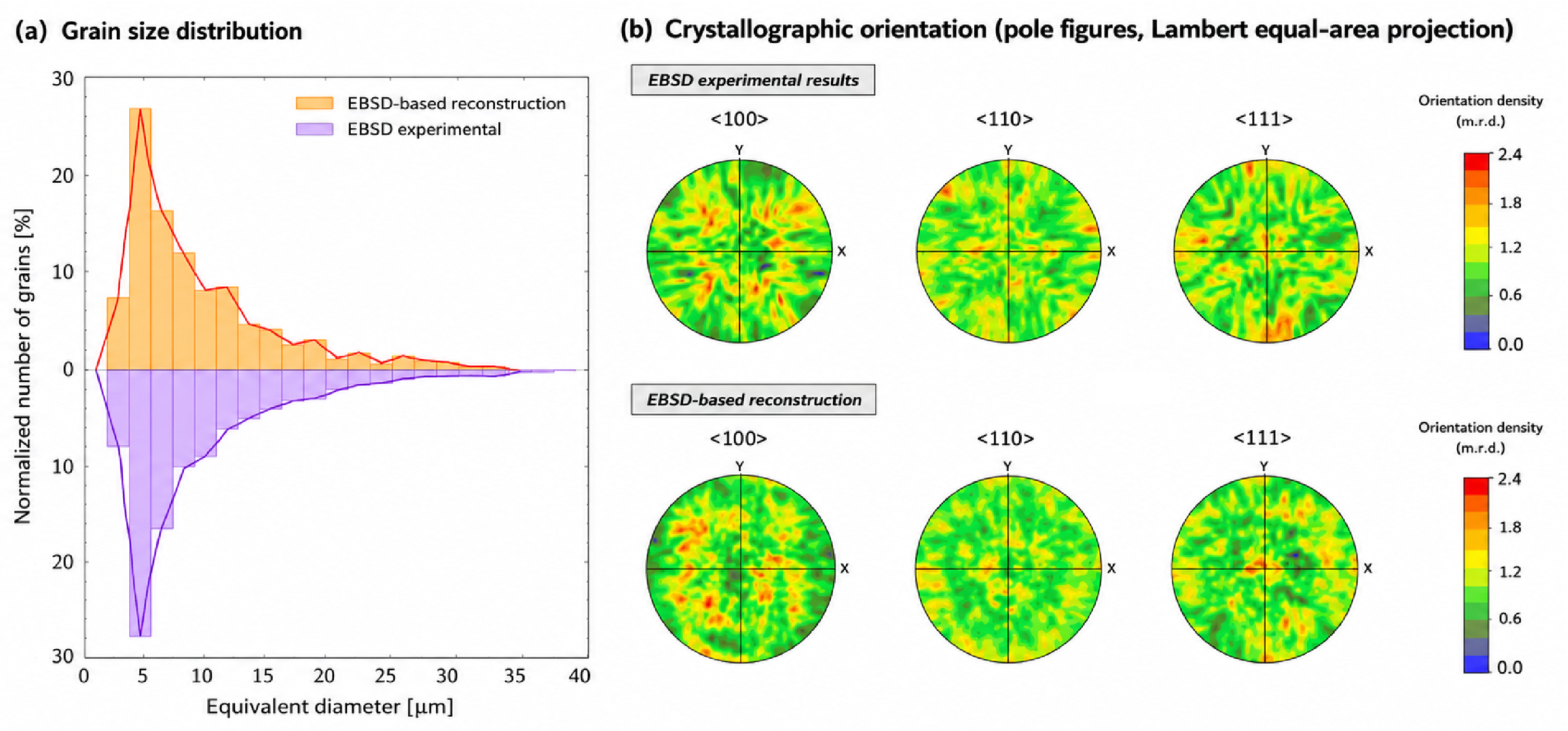}
		\caption{Verification of microstructure reconstruction based on: (a) grain size distribution and (b) crystallographic orientation distribution represented by contoured pole figures ($\langle 100 \rangle$,$\langle 110 \rangle$,$\langle 111 \rangle$) derived from EBSD measurements and the EBSD-based reconstructed microstructure. Pole figures are shown as orientation density distributions in a Lambert equal-area projection.}
		\label{fig:Ver}
\end{figure}

Figure~\ref{fig:EBSDmaps} presents the EBSD results of the polycrystalline NiAl sample, including the as-received inverse pole figure (IPF) orientation map and the corresponding reconstructed grain-size representation after smoothing and segmentation. The IPF map (Fig.~\ref{fig:EBSDmaps}a) reveals a heterogeneous polycrystalline microstructure composed of grains with a broad size distribution and varying crystallographic orientations. The observed color variations in the IPF map indicate differences in orientation between neighboring grains, indicating the absence of a pronounced preferred orientation at the microscale. The processed map (Fig.~\ref{fig:EBSDmaps}b) enables quantitative visualization of the spatial grain size distribution by separating individual grains and assigning colors according to equivalent grain diameter. 
The grain size distribution appears broad and markedly heterogeneous, ranging from fine grains of a few micrometers to coarse grains exceeding 130\,$\mu$m in equivalent diameter. Such a conclusion can be confirmed by Figure~\ref{fig:Ver}a, which presents the histogram obtained from segmented data. The distribution is strongly skewed toward smaller grain diameters, with the highest fraction of grains concentrated at equivalent diameters of approximately 4–6\,$\mu$m, indicating a predominance of fine grains within the microstructure. The frequency of grains decreases progressively with increasing equivalent diameter, forming a long tail extending to approximately 35–38\,$\mu$m, which confirms the presence of a limited population of coarser grains. As shown, the segmentation and smoothing data allowed the extraction of statistical descriptors required for microstructural quantification and reconstruction. 

The main requirements for generating a synthetic polycrystalline NiAl sample were to account for the actual grain-size and orientation distributions, thereby improving the comparison between experimental and numerical elastic-constant results. As shown in Fig.~\ref{fig:Ver}b, the EBSD-based reconstruction preserves the overall orientation distribution observed experimentally. The pole figures evaluated for the EBSD-reconstructed microstructure show good agreement with the experimental EBSD measurements, confirming that the material exhibits neither crystallographic texture nor pronounced preferential orientation. No significant texture distortion was introduced during reconstruction, as confirmed by the qualitative similarity of the $\langle 100 \rangle$, $\langle 110 \rangle$ and $\langle 111 \rangle$ pole figures.

Finally, the reconstructed microstructure accurately reproduces the characteristic shape of the experimental distribution (Fig.~\ref{fig:Ver}a), including the location of the dominant peak, the decay tendency at intermediate grain sizes, and the low-frequency occurrence of large grains. It should be emphasized that the largest grains in the intermetallic sample (equivalent diameter $> 40\,\mu$m) were excluded from the reconstruction due to the limited size of the numerical domain. Since a real experimental specimen may contain billions of grains, explicitly representing every grain in a finite element framework remains computationally impractical with currently available resources. Therefore, computational efficiency was improved by reducing the model size while preserving the original grain-size distribution and packing configuration. Despite minor differences between the grain characteristics of the experimental and reconstructed polycrystals, the average grain sizes remained comparable, amounting to approximately $\sim 12\,\mu$m and $\sim 10\,\mu$m, respectively. The reduced numerical domain was assumed to provide a statistically representative description of the real NiAl microstructure.

\begin{figure}[H]
    \centering
    \includegraphics[width=1\linewidth]{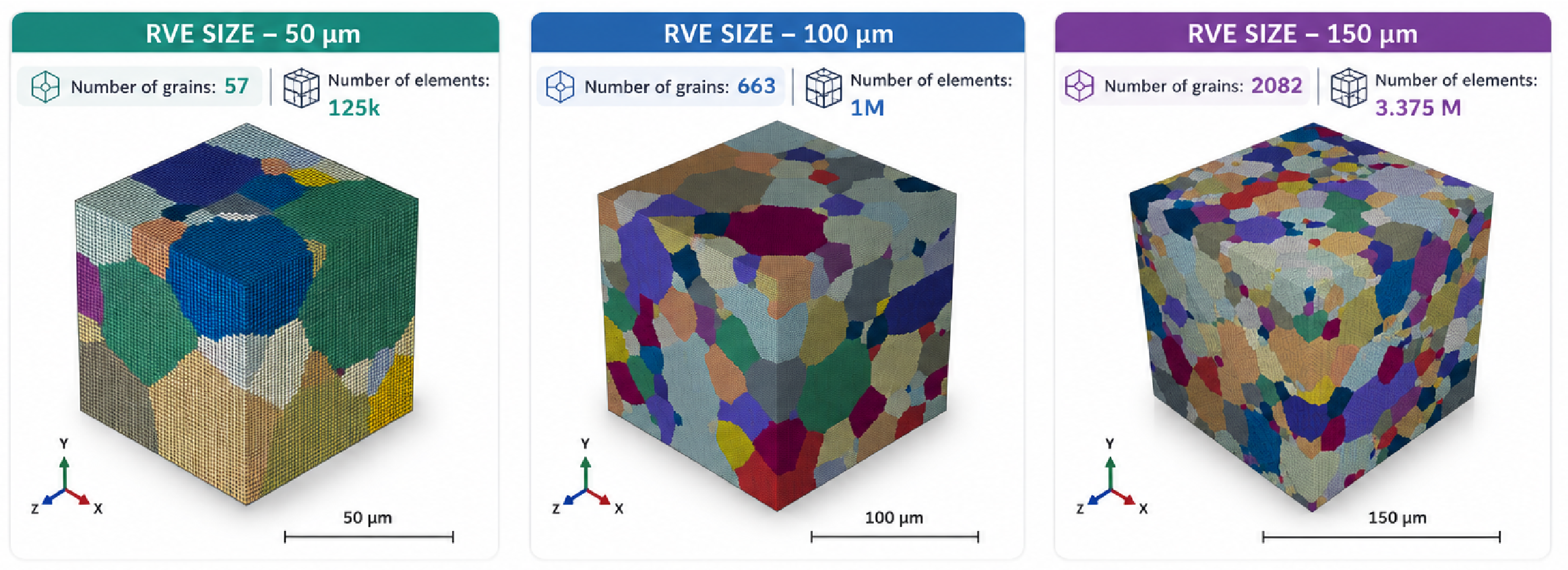}
    \caption{Generated EBSD-based FEM models: RVE comparison.}
    \label{fig:FEMmodels}
\end{figure}

Figure~\ref{fig:FEMmodels} presents EBSD-reconstructed representative volume elements (RVEs) with characteristic dimensions of 50\,$\mu$m, 100\,$\mu$m, and 150\,$\mu$m, illustrating the progressive increase in microstructural complexity as model size increases. As the RVE dimension increases, the number of explicitly represented grains rises significantly from 57 to 2082, accompanied by a corresponding increase in mesh density from 125$\times$10$^3$ to 3.375$\times$10$^6$ finite elements, enabling a more statistically representative reconstruction of the polycrystalline body. Here, the main assumption underlying the generation of the three-dimensional EBSD-based geometrical model is the isotropic character of the material, which is supported by the experimentally observed nearly random grain-orientation distribution, as confirmed by pole figure analysis. The smallest RVE (50\,$\mu$m) contains only a limited number of grains and therefore exhibits a simplified microstructural topology, which may not fully capture the statistical heterogeneity of the experimental material. In contrast, the larger RVEs (100\,$\mu$m and 150\,$\mu$m) provide a substantially improved representation of grain morphology, spatial distribution, and grain interaction effects, approaching a statistically representative description of the NiAl microstructure. We note that a similar number of finite elements per grain are used to mesh the RVEs of 100 and 150~$\mu m$ size. Increasing the RVE size enhances microstructural representativeness but also substantially increases computational cost due to the substantially larger number of finite elements and degrees of freedom \cite{liu_advances_2026}. Therefore, selecting an appropriate RVE size represents a compromise between computational efficiency and the ability to accurately reproduce the experimentally observed microstructural characteristics of the polycrystalline system.

    \subsection{Finite element simulations of EBSD-based polycrystal}
    \label{sec:FesoEp}

\begin{figure}[!ht]
    \centering
    \includegraphics[width=0.8\linewidth]{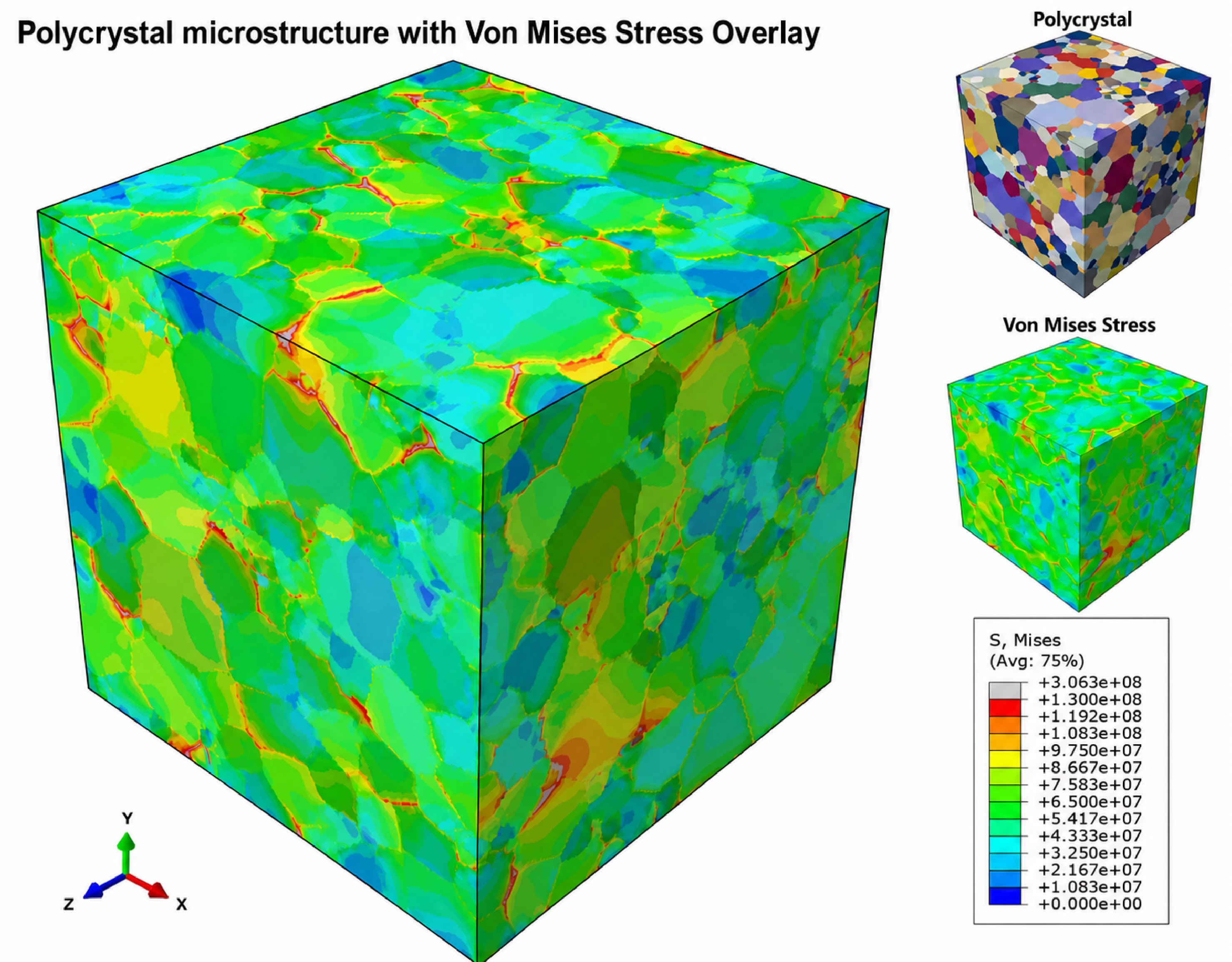}
    \caption{Representative visualization of the reconstructed polycrystalline microstructure with the corresponding von Mises stress distribution obtained from finite element simulations.}
    \label{fig:VonMises}
\end{figure}

Generated samples were next applied to the FE simulations via the procedure presented in Sec.~\ref{sec:FEM}. Figure~\ref{fig:VonMises} shows a representative three-dimensional visualization of the reconstructed polycrystalline microstructure (RVE - 150\,$\mu$m)  combined with the corresponding von Mises stress distribution obtained from finite element simulations. The central cube presents a transparent overlay of the reconstructed polycrystalline morphology and the von Mises stress field, enabling simultaneous observation of grain topology and local stress localization. 

The von Mises stress distribution reveals a strongly heterogeneous mechanical response within the reconstructed polycrystalline volume, characterized by pronounced local stress concentrations and regions of relatively low equivalent stress. The highest von Mises stress values are localized in narrow regions primarily situated near grain boundaries, triple junctions, and geometrically constrained microstructural features. These localized maxima indicate preferential sites of stress accumulation caused by incompatibilities in local deformation between neighboring grains and the associated redistribution of internal stresses. The minimum von Mises stress values, approaching 0\,Pa, are observed in localized regions exhibiting low deformation incompatibility and reduced mechanical constraint. These low-stress domains are predominantly located within larger grains or in microstructural regions where local stress transfer is comparatively limited. Such areas represent mechanically relaxed zones, where load redistribution leads to lower equivalent stress magnitudes.

Next, the local microscopic stress of each grain has been correlated with the loading direction and grain orientation in the reconstructed sample. The grain-averaged von Mises stress, $\bar{\sigma}_{\mathrm{VM}}^{(g)}$, was determined by volume-averaging the element-wise von Mises stresses within a given grain according to: 

\begin{equation}
\bar{\sigma}_{\mathrm{VM}}^{(g)}
=
\frac{\displaystyle\sum_{i=1}^{N_g}
\sigma_{\mathrm{VM}}^{(i)} V_i}
{\displaystyle\sum_{i=1}^{N_g} V_i}
\label{eq:grain_vm}
\end{equation}

where $\sigma_{\mathrm{VM}}^{(i)}$ denotes the von Mises stress of the $i$-th finite element, $V_i$ is the corresponding finite element volume, and $N_g$ represents the total number of finite elements belonging to grain $g$. This averaging procedure provides a representative measure of the local stress state within each grain while minimizing the influence of mesh discretization.

The crystallographic orientation of each grain $g$ was represented by the orientation matrix $\mathbf{R}_g$, obtained from the corresponding Euler angles $\bfig=(\phi_1,\phi_2,\phi_3)_g$ using the Bunge convention \cite{Maitland2007, depriester_computing_2018} \comm{\cite{maitland_electron_2015, depriester_computing_2018}}. The equivalent crystallographic directions belonging to the $\langle100\rangle$ family were transformed from the crystal coordinate system to the global (sample) coordinate system according to

\begin{equation}
\mathbf{d}_i^{\,s} = \mathbf{R}_g \mathbf{d}_i^{\,c},
\end{equation}

where $\mathbf{d}_i^{\,c}$ and $\mathbf{d}_i^{\,s}$ are unit vectors representing the $i$-th crystallographic direction in the crystal and sample reference frames, respectively, and $\mathbf{R}_g$ is the orientation matrix of grain $g$.

The angular deviation between each transformed crystallographic direction and the loading direction $\mathbf{l}^{\,s}$ was calculated as

\begin{equation}
\theta_i =
\cos^{-1}
\left(
\frac{
\left| \mathbf{d}_i^{\,s}\cdot\mathbf{l}^{\,s} \right|
}{
\left\|\mathbf{d}_i^{\,s}\right\|
\left\|\mathbf{l}^{\,s}\right\|
}
\right),
\end{equation}

where the absolute value accounts for the crystallographic equivalence of opposite directions. Finally, the grain misorientation angle was defined as the minimum angular deviation between the loading direction and the equivalent members of the $\langle100\rangle$ direction family,

\begin{equation}
\theta_{\langle100\rangle}
=
\min
\left(
\theta_{100},
\theta_{010},
\theta_{001}
\right).
\end{equation}

\begin{figure}[!ht]
    \centering
    \includegraphics[width=1\linewidth]{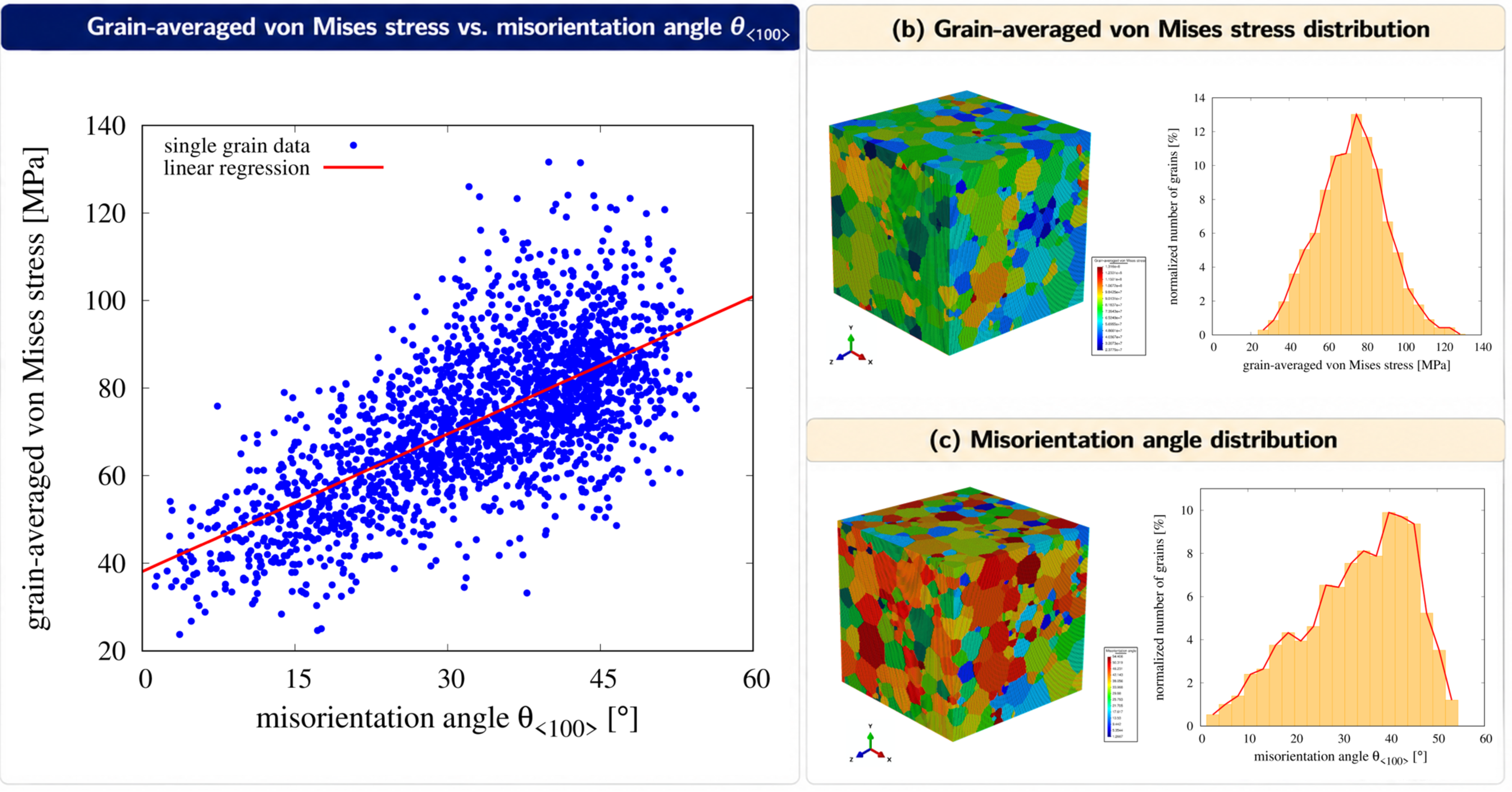}
    \caption{Influence of crystallographic orientation on stress partitioning in the reconstructed polycrystal: (a) Grain-averaged von Mises stress as a function of the misorientation angle $\theta_{\langle100\rangle}$ between the loading direction and the nearest equivalent $\langle100\rangle$ crystal direction, (b) Distribution of grain-averaged von Mises stress within the representative volume element (RVE), (c) Distribution of the misorientation angle $\theta_{\langle100\rangle}$.}
    \label{fig:MisAngle}
\end{figure}

The obtained parameter $\theta_{\langle100\rangle}$ provides a quantitative measure of the alignment between the grain crystallography and the macroscopic loading direction, ranging from $0^\circ$ for perfect alignment with a $\langle100\rangle$ direction to $54.74^\circ$ for orientations closest to the $\langle111\rangle$ direction. It should be noted that the $\langle100\rangle$ directions correspond to the lowest directional Young's modulus of crystalline NiAl, whereas the $\langle111\rangle$ directions---to the highest one (Sec.~\ref{sec:Mds}). The influence of crystallographic orientation on stress distribution within the reconstructed polycrystal is illustrated in Fig.~\ref{fig:MisAngle}. 

A positive correlation is observed between the grain-averaged von Mises stress and the misorientation angle $\theta_{\langle100\rangle}$, indicating that grains whose nearest $\langle100\rangle$ crystallographic direction is less aligned with the loading axis tend to experience higher stress levels. The linear regression trend confirms that the average stress carried by individual grains increases with increasing angular deviation from the preferred loading orientation. The three-dimensional distribution of grain-averaged von Mises stress reveals pronounced heterogeneity throughout the representative volume element (RVE), reflecting the anisotropic mechanical response of neighboring grains (Fig.~\ref{fig:MisAngle}b). Regions characterized by elevated stress values are associated with specific orientation arrangements, demonstrating the strong influence of local crystallographic orientation on stress redistribution. The histogram of grain-averaged stress values exhibits a nearly unimodal distribution centered around intermediate stress levels, whereas the misorientation angle distribution is skewed toward larger angular deviations (Fig.~\ref{fig:MisAngle}c).

\subsection{Numerical and theoretical predictions of elastic constants}
\label{sec:Natpoec}

Since the effective elastic constants of polycrystalline materials depend on the statistical representation of the microstructure, determining an appropriate RVE size is essential. In particular, the RVE must contain a sufficient number of grains to ensure convergence of the effective Young's modulus, Poisson's ratio, and the associated anisotropy measures. Otherwise, local microstructural heterogeneities may lead to non-representative predictions of the macroscopic elastic response. Table~\ref{tab:RVE_anisotropy} summarizes the influence of RVE size on the effective elastic properties and their variability among individual realizations. 

\begin{table}[htbp]
\centering
\caption{Influence of RVE size on the effective Young's modulus, Poisson's ratio, and its deviation from isotropy (FEM calculations).}
\label{tab:RVE_anisotropy}
\begin{tabular}{cccccccccc}
\hline
\textbf{RVE size} &
\textbf{Samples / } &
\boldmath$\overline{E}$ &
\boldmath$SD(E)$ &
\boldmath$CoV(E)$ &
\boldmath$\overline{A_E}$ &
\boldmath$SD(A_E)$ &
\boldmath$\overline{\nu}$ &
\boldmath$SD({\nu})$ &
\boldmath$\overline{A_{\nu}}$ \\
\textbf{($\mu$m)} &
\textbf{evaluations} &
\textbf{(GPa)} &
\textbf{(GPa)} &
\textbf{(\%)} &
\textbf{(\%)} &
\textbf{(\%)} &
\textbf{(-)} &
\textbf{(-)} & 
\textbf{(\%)} \\
\hline
50  & 9 / 54 & 187.3 & 21.0 & 11.2 & 18.0 & 7.4 & 0.303 & 0.0241 & 13.8 \\
100 & 6 / 36 & 181.5 & 5.9 & 3.2 & 6.3 & 2.0 & 0.310 & 0.0068 & 4.4 \\
150 & 4 / 24 & 182.2 & 2.9 & 1.6 & 3.2 & 1.0 & 0.309 & 0.0034 & 2.3 \\
\hline
\end{tabular}
\end{table}

The average Young's modulus remained relatively stable with increasing RVE size, varying only from 187.3\,GPa for the 50\,$\mu$m RVEs (for over 50 realizations/simulations) to approximately 182\,GPa for the larger models, indicating convergence of the effective stiffness. In contrast, the standard deviation and coefficient of variation of Young's modulus decreased substantially with increasing RVE size, with $CoV(E)$ reducing from 11.2\% for 50\,$\mu$m RVEs to only 1.6\% for 150\,$\mu$m RVEs. A similar trend was observed for the anisotropy index $A_E$, whose mean value decreased from 18.0\% to 3.2\%, demonstrating a progressive reduction in elastic anisotropy and improved representativeness of larger RVEs. The effective Poisson's ratio remained nearly constant ($\bar{\nu} \approx 0.31$) regardless of RVE size; however, its standard deviation and anisotropy measure $A_{\nu}$ also decreased significantly as the RVE dimensions increased. These findings indicate that both larger RVEs with sizes 100 and 150\,$\mu$m, provide a more statistically representative description of the polycrystalline aggregate, yielding both stabilized effective elastic properties and reduced sensitivity to microstructural orientation effects.

\begin{figure}[!ht]
    \centering
    \includegraphics[width=1\linewidth]{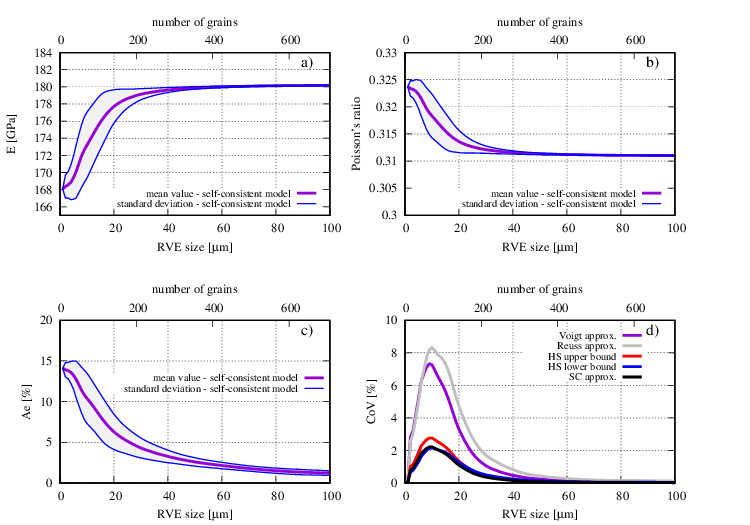}
    \caption{Dependence of the effective elastic response on the RVE size: (a) Young's modulus, (b) Poisson's ratio, (c) elastic anisotropy index and (d) coefficient of variation (CoV) for different models.}
    \label{fig:RVE_multiplot}
\end{figure}

The high computational efficiency of the analytical homogenization methods made it possible to perform a more extensive analysis of the representative volume element (RVE) size. Figure~\ref{fig:RVE_multiplot} demonstrates the convergence of the effective elastic properties (evaluated by the self-consistent model) with increasing RVE size. As the RVE size increases, both the effective Young's modulus and Poisson's ratio rapidly converge toward stable values of approximately 180\,GPa and 0.311, respectively, indicating that the reconstructed microstructure becomes statistically representative. The largest reduction in variability occurs for RVE sizes below approximately 40\,$\mu$m, whereas further enlargement produces only marginal changes in the predicted properties. Moreover, the self-consistent and Hashin--Shtrikman models exhibit consistently lower scatter than the Voigt and Reuss estimates, suggesting that they provide more robust predictions of the effective elastic response of polycrystalline NiAl.

Finally, the numerical, analytical and experimental results of the effective Young's modulus and Poisson's ratio were compared. The experimental elastic properties used to validate the proposed multiscale framework were taken from the ultrasonic measurements reported in \cite{chmielewski_sintering_2014}. The experimental Young's modulus and Poisson's ratio were determined using the pulse-echo ultrasonic technique on sintered polycrystalline NiAl samples, providing a reliable reference for comparison with the numerical and theoretical predictions. 

The comparison presented in Table~\ref{tab:comparison} demonstrates that the reconstructed polycrystalline FEM model accurately captures the effective elastic response of the material. The predicted Young's modulus and Poisson's ratio differ from the experimental values by only -0.5\% and -2.8\%, respectively, indicating excellent agreement. While the classical Voigt and Reuss schemes establish theoretical upper and lower bounds for the elastic properties, their deviations from experimental measurements can reach up to 30.8\%, highlighting the limitations of these simplified assumptions for the investigated microstructure. The Hashin--Shtrikman upper bound and self-consistent approximation provide substantially improved estimates, with deviations below 2\% for the effective Young's modulus. 

\begin{table}[htbp]
\centering
\caption{Comparison of the effective Young's modulus and Poisson's ratio obtained from experiments (ultrasonic measurement) \cite{chmielewski_sintering_2014}, FEM simulations, and theoretical homogenization schemes. Relative deviations from experiment, $\Delta E$ and $\Delta \nu$, are calculated with respect to the experimentally measured values.}
\label{tab:comparison}
\begin{tabular}{lcccccc}
\hline
\textbf{Method} &
\boldmath$\overline{E}$ \textbf{(GPa)} &
\textbf{SD(\boldmath$E$)} \textbf{(GPa)} &
\boldmath$\Delta E$ \textbf{(\%)} &
\boldmath$\overline{\nu}$ \textbf{(-)} &
\textbf{SD(\boldmath$\nu$)} \textbf{(-)} &
\boldmath$\Delta \nu$ \textbf{(\%)}\\
\hline

Experiment
& 183.1 & 1.5 & -- & 0.318 & 0.0025 & -- \\

FEM
& 182.2 & 2.9 & -0.5 & 0.309 & 0.0034 & -2.8 \\

Voigt bound
& 210.8 & 0.043 & +15.1 & 0.279 & 0.00005 & -12.3 \\

Reuss bound
& 126.7 & 0.048 & -30.8 & 0.367 & 0.00005 & +15.4 \\

Hashin--Shtrikman upper approx.
& 183.2 & 0.015 & +0.1 & 0.308 & 0.00002 & -3.1 \\

Hashin--Shtrikman lower approx.
& 155.9 & 0.018 & -14.9 & 0.336 & 0.00002 & +5.7 \\

Self-consistent model
& 180.2 & 0.013 & -1.6 & 0.311 & 0.00001 & -2.2 \\

\hline
\end{tabular}
\end{table}

\section{Discussion}

The present multiscale framework demonstrates that advanced analytical homogenization methods, particularly the Hashin--Shtrikman upper bound and the self-consistent model, provide accurate predictions when combined with atomistically derived elastic constants and EBSD-reconstructed microstructures, while remaining computationally efficient. The self-consistent model represents each anisotropic grain as an inclusion embedded in an equivalent homogeneous medium whose elastic properties are equal to the unknown effective properties of the polycrystal \cite{molinari_self-consistent_1997}. This assumption naturally accounts for the average mechanical interaction among neighboring grains and provides an accurate representation of randomly oriented polycrystalline aggregates \cite{li_constitutive_2022}. Similarly, the Hashin--Shtrikman upper bound is derived from variational principles and incorporates information on the phase morphology through an optimal reference medium, yielding one of the tightest theoretical bounds on the effective elastic properties \cite{kube_bounds_2016}. For dense, fully bonded polycrystalline materials with nearly random crystallographic orientation, the assumptions of both models closely resemble the actual mechanical response, thereby explaining their excellent agreement with finite element simulations based on EBSD-reconstructed microstructures and experimental measurements.

Indicating close agreement with the measured effective modulus with experimental results, the FEM approach extends analytical homogenization by explicitly accounting for grain interactions and local stress–strain heterogeneity \cite{zhang_effects_2022}. It has been demonstrated that atomistic-based multiscale frameworks offer the fundamental advantage of determining constitutive parameters directly from atomic interactions, thereby reducing the reliance on empirically fitted material parameters and establishing a rigorous physical link between the atomistic and macroscopic scales \cite{huang_multiscale_2026}. However, the accuracy of the predicted effective properties depends directly on the quality of the atomistic input. Fri\'ak et al. \cite{friak_error_2008} showed that uncertainties in single-crystal elastic constants can propagate across the scales and influence the homogenized elastic response. {\color{black}
On the one hand, experimental single-crystal elastic constants could also serve as input to the micromechanical model \cite{Rusovic1977}. On the other hand, the experimental datasets compiled by Ponomareva et al. \cite{Ponomareva2012} show appreciable variability, with reported single-crystal elastic constants ranging from 197--212\,GPa for $C_{11}$, 119--143\,GPa for $C_{12}$, and 110--117\,GPa for $C_{44}$. 

The adopted EAM2009 potential combines fitting to experimental single-crystal elastic constants of B2-NiAl with broad transferability across the binary Ni--Al system, reproducing several properties outside its fitting database, including thermal expansion of NiAl and elastic constants of Ni$_3$Al \cite{Purja2009}. Here, atomistic calculations were performed using the same potential and simulation procedure for three orientations obtained by rotating the entire atomistic crystal model. Agreement between the directly
calculated stiffness tensors and those obtained by mathematically rotating the reference tensor, together with the invariance of the Kelvin moduli, provides numerical verification of the orientation-dependent constitutive input before its transfer to the finite element model, consistent with the importance of numerical verification in atomistic elasticity calculations \cite{Hale2018}.
The main components of the NiAl stiffness tensor calculated in the crystallographic reference frame (oriented in the \textbf{x}=[100], \textbf{y}=[010], and \textbf{z}=[001] directions) are consistent with the corresponding literature data \cite{fu_elastic_2013, Mishin2002, Papanicolaou2003, Faraoun2005}; however, the higher Zener ratio predicted by EAM2009 relative to the experimental benchmark \cite{LaRosa2025} remains a limitation when interpreting the predicted directional stiffness and local stress distributions. 
}

Despite the satisfactory agreement obtained in the validation, the proposed multiscale framework may be further improved by incorporating an explicit description of grain-boundary regions. A suitable extension is provided by the core–shell concept, which models each grain as an anisotropic crystalline core surrounded by an isotropic grain-boundary shell, thereby accounting for grain-boundary interactions while maintaining the computational efficiency of mean-field homogenization \cite{kowalczyk-gajewska_elastic_2020, kowalczyk-gajewska_effective_2019}. Alternatively, a widely adopted strategy for explicitly modeling grain-boundary interactions consists of inserting cohesive-zone elements along grain interfaces within a conforming finite element mesh \cite{kowalski_finite_2016}. Such an approach enables the simulation of displacement discontinuities, intergranular debonding, and crack initiation and propagation \cite{demir_finite_2026} while preserving the anisotropic elastic response of individual grains. An alternative approach treats grain boundaries as a distinct finite-thickness phase discretized by dedicated finite elements rather than zero-thickness cohesive interfaces \cite{majewski2017micromechanical}. In this framework, the grain-boundary region is assigned its own constitutive behavior, with elastic and fracture properties obtained directly from atomistic simulations of amorphous structures representing the disordered grain-boundary material \cite{mazdziarz_atomistic_2024}. Finally, the explicit representation of grain-boundary mechanics can be modeled through dedicated continuum formulations, including enriched finite element methods with displacement discontinuities \cite{simone_generalized_2006} and analytical or semi-analytical models for predicting local grain-boundary stresses \cite{el_shawish_grain_2023}.

Although the aforementioned approaches offer promising directions for future development, the present multiscale framework, derived from EBSD data, adequately reproduces the crystallographic orientation distribution and grain morphology that govern the macroscopic elastic behavior. A potential extension of the proposed framework is the incorporation of three-dimensional EBSD reconstructions into the finite element model \cite{zaafarani_three-dimensional_2006}. By reconstructing the complete grain network, including grain connectivity and three-dimensional grain morphology \cite{swanson_quantitative_2026}, such models can provide a more accurate description of local mechanical interactions than their two-dimensional counterparts. However, because the effective elastic properties predicted in the present study are in excellent agreement with experimental measurements, the adopted two-dimensional EBSD reconstruction appears sufficient for capturing the dominant microstructural features governing the macroscopic elastic response, while maintaining substantially lower experimental and computational costs.

{\color{black}
While the present study considers a relatively simple, dense NiAl polycrystal, the framework can be extended to more complex microstructures, including porous materials \cite{nosewicz_combined_2022}, multiphase composites \cite{Gabel2021}, superalloys \cite{Gabel2022}, and coatings. Such extensions would require explicit representation of pores or constituent phases, appropriate crystallographic orientation
distributions, and constitutive properties for each phase. Another relevant application concerns materials with pronounced crystallographic texture produced by severe plastic deformation, including high-pressure torsion \cite{Jahedi2014}. Incorporating their measured orientation distributions and retaining the full effective stiffness tensor would enable assessment of texture-induced elastic anisotropy and its influence on local stress partitioning, quantities relevant to the micromechanical behavior of anisotropic polycrystals \cite{Brenner2009}.}

	\section{Conclusions}
	\label{sec:Conc}
	
	The presented work can be summarized in the following remarks:
	
\begin{itemize}

\item A multiscale framework integrating molecular dynamics, EBSD-based microstructure reconstruction, finite element modeling, and analytical homogenization schemes was successfully developed to determine the effective elastic properties of polycrystalline NiAl. The proposed methodology enables a physically consistent transfer of material information from the atomistic scale to the polycrystalline microscale while preserving realistic microstructural characteristics.

\item Molecular dynamics simulations provided the complete stiffness tensor of B2-NiAl and confirmed its pronounced cubic elastic anisotropy. The directional elastic response exhibited the maximum Young's modulus along the $\langle111\rangle$ directions and the minimum stiffness along the $\langle100\rangle$ directions, while the directional average Poisson's ratio also showed substantial anisotropy, highlighting the strong influence of crystallographic orientation on both stiffness and transverse deformation.

\item The EBSD-based reconstruction methodology successfully reproduced the essential characteristics of the experimental microstructure. The generated three-dimensional representative volume elements preserved the experimentally observed nearly random grain-orientation distribution and provided the actual grain-size distribution of the polycrystalline NiAl material.

\item The representative volume element study confirmed that the effective elastic response converged as the model size increased. Based on analytical homogenization results, the largest reduction in statistical scatter occurs for RVEs smaller than approximately $40$--$50\,\mu$m. In turn, the FEM calculations demonstrate that RVEs of $100$--$150\,\mu$m are sufficient to accurately reproduce the effective Young's modulus, Poisson's ratio, and macroscopic isotropic behavior with low statistical uncertainty.

\item Finite element simulations revealed a highly heterogeneous stress distribution within the reconstructed polycrystal, with stress concentrations localized primarily near grain boundaries and triple junctions. Furthermore, a positive correlation was identified between the grain-averaged von Mises stress and the misorientation angle relative to the nearest $\langle100\rangle$ crystallographic direction characterized by the lowest value of directional Young's modulus, demonstrating the important role of grain orientation in stress partitioning.

\item The largest reconstructed RVE ($150\,\mu$m, 2082 grains, 3.375$\times$10$^6$ finite elements) reproduced the experimentally measured elastic properties with satisfactory accuracy, yielding deviations of only $-0.5\%$ and $-2.8\%$ for Young's modulus and Poisson's ratio, respectively. Among the analytical homogenization schemes, the Hashin--Shtrikman upper bound and self-consistent approximation provided the closest agreement with experiments, whereas the Voigt and Reuss approaches established the theoretical upper and lower bounds of the effective elastic response.

\end{itemize}

\comm{	\section*{CRediT authorship contribution statement}
	\noindent \textbf{Szymon Nosewicz} Conceptualization, Funding acquisition, Methodology, Visualization, Writing – original draft, Writing – review \& editing.\\ 
	\textbf{Marcin Ma\'zdziarz:} Conceptualization, Methodology, Resources, Software, Visualization, Writing – original draft.\\
	\textbf{Grzegorz Jurczak} Conceptualization, Methodology, Visualization, Writing – original draft, Writing – review \& editing, Funding acquisition.}
	
	\section*{Declaration of generative AI and AI-assisted technologies in the writing process}
	During the preparation of this work, the authors used Grammarly and ChatGPT (by OpenAI) to correct spelling and English grammar and to improve the clarity of the images. After using this tool/service, the authors reviewed and edited the content as needed and assume full responsibility for the publication's content.
	
	\section*{Declaration of Competing Interest}
	The authors declare that they have no known competing financial interests or personal relationships that could have appeared to influence the work reported in this paper.
	
	\section*{ACKNOWLEDGMENTS}
	The authors would like to acknowledge the financial support of the National Science Centre, Poland, under Grant Agreement No. OPUS 2020/37/B/ST8/03907 for the project “Multiscale investigation of deformation and damage behavior of novel hybrid metal matrix composites. Experimental studies and numerical modeling”.
	Additional support for the work was provided by the GRAFEN computing cluster at Biocentrum Ochota, the Interdisciplinary Centre for Mathematical and Computational Modelling of Warsaw University (ICM UW) and Pozna\'n Supercomputing and Networking Center (PSNC).
	
\bibliographystyle{elsarticle-num}
\bibliography{referencesAll}

\end{document}